\documentclass[%
  reprint,
]{revtex4-2}

\usepackage{amsmath}
\usepackage{amsthm}
\usepackage{amssymb}
\usepackage{bm,bbm}

\usepackage[T1]{fontenc}
\usepackage[sc,osf]{mathpazo}
\usepackage{dsfont}

\usepackage{physics}

\usepackage{graphicx}
\usepackage{xcolor}
\usepackage[bookmarks=false]{hyperref}
\hypersetup{colorlinks=true,citecolor=blue,linkcolor=red,%
urlcolor=blue,pdfstartview=FitH,bookmarksopen=true}

\newtheorem{theorem}{Theorem}

\newtheorem{lemmab}{Lemma}[section]

\newtheorem{observation}{Observation}
\newtheorem{theoremb}{Theorem}[section]

\newcommand{\maxover}[1][]{\underset{#1}{\mathrm{max}}}
\newcommand{\minover}[1][]{\underset{#1}{\mathrm{min}}}
\newcommand{\subto}{\mathrm{~s.t.}}

\newcommand{\I}{\mathds{1}}
\newcommand{\dC}{\mathds{C}}
\newcommand{\dR}{\mathds{R}}
\newcommand{\cH}{\mathcal{H}}
\newcommand{\cD}{\mathcal{D}}
\newcommand{\cE}{\mathcal{E}}

\newcommand{\Id}{\mathrm{id}}

\newcommand{\tL}{\tilde{\Lambda}}
\newcommand{\tD}{\tilde{\mathcal{D}}}

\newcommand{\stateset}{\{\rho_1,\rho_2,\dots,\rho_m\}}

\begin{document}
\title{Virtual Cloning of Quantum States}
\author{Zhi-Hao Bi}
\affiliation{Department of Physics, Shandong University, Jinan 250100, China}
\author{Jing-Tao Qiu}
\affiliation{Department of Physics, Shandong University, Jinan 250100, China}
\author{Xiao-Dong Yu}
\email{yuxiaodong@sdu.edu.cn}
\affiliation{Department of Physics, Shandong University, Jinan 250100, China}

\date{\today}

\begin{abstract}
The inherent limitations of physical processes prevent the copying of
arbitrary quantum states. Furthermore, even if we only aim to clone two
distinct quantum states, it remains impossible unless they are mutually
orthogonal. To overcome this limitation, we propose a virtual-cloning protocol
that bypasses the restrictions imposed by the quantum no-cloning theorem. 
Specifically, we begin by outlining the general framework for virtual cloning
and deriving a necessary and sufficient criterion for the existence
of a virtual operation capable of simultaneously cloning a set of states.
Subsequently, through an analysis of the simulation cost of the virtual-cloning
process, we demonstrate that the problem of identifying an optimal virtual-cloning 
protocol can be cast as a semidefinite programming problem.
Finally, we establish a connection between virtual cloning and state
discrimination, from which universal bounds on the optimal cloning cost are
derived.
\end{abstract}
\maketitle

\section{Introduction}

In classical information processing, copying or cloning information is a
fundamental operation. Classical bits can be duplicated without any inherent
limitations, allowing for straightforward replication and transmission of data.
This ease of copying is a cornerstone of classical computing systems. In
contrast, quantum information processing operates under different principles.
The no-cloning theorem, a foundational result in quantum information theory,
states that it is impossible to create an exact copy of an arbitrary unknown
quantum state \cite{nocloningWootters1982,nosuperluminalDIEKS1982}.
The no-cloning theorem plays a dual role in quantum information
\cite{CloningRevModPhys2005,CloningreviewFAN2015}.
It preserves causality and prevents superluminal communication, ensuring that
quantum entanglement does not violate special relativity. 
Simultaneously, it underpins the security of quantum communication protocols by
preventing eavesdroppers from cloning quantum states and extracting information 
without introducing detectable disturbances. 
This guarantees the privacy of transmitted data and enables secure quantum
cryptographic protocols, providing a level of security unmatched by
classical methods due to the impossibility of cloning quantum states.

While perfect cloning of arbitrary unknown quantum states is impossible, it is
possible to approximate cloning with optimal fidelity or to achieve perfect cloning
with the highest probability. Various quantum cloning machines have been
developed to support different quantum information protocols.
Notable examples include the symmetric universal quantum cloning machine 
\cite{CloningOptimal1-2UQCM,CloningOptimalN-MUQCM,Cloning1-2UQCMd-dim},
the asymmetric universal quantum cloning machine 
\cite{CloningAsymmetricNiu1998,CloningAsymmetricPauli2000},
the probabilistic quantum cloning machine
\cite{CloningProbabilisticDuan,CloningProbabilisticPati},
and the phase-covariant quantum cloning machine
\cite{CloningPhaseBru2000,CloningPhaseFan2001}.
These quantum cloning machines play a crucial role
in the security analysis of quantum key distribution 
\cite{CloningPhaseNiu,EavesdroppingBruss2002,EavesdroppingCerf2002},
quantum state estimation
\cite{CloningOptimalN-MUQCM,StateEstiBru1998},
quantum measurement compatibility
\cite{MeasuCompaBrougham2006,MeasuCompaSagawa2008,MeasuCompaThekk2017},
and the foundations of quantum mechanics 
\cite{Cloning-EntanglementBruss2003,
Cloning-EntanglementLamoureux2004,Cloning-EntanglementWeedbrook2008}.

Despite all these efforts, the fundamental limitation on cloning
nonorthogonal states still persists due to constraints imposed by
standard quantum operations \cite{no-cloning2YUEN1986,nobroadcasting}. 
In this work, we explore  quantum cloning from a different perspective. We
expand the set of allowable operations to include virtual quantum operations.
These operations, also referred to as Hermitian-preserving, trace-preserving
(HPTP) maps, are a class of linear maps that transform Hermitian operators into
Hermitian operators while preserving the trace. Virtual quantum operations have
been widely used in various quantum information tasks, including error mitigation 
\cite{ErrorMitigationShortDepthCircuits,Jiangphysicalimplement,
ErrorMitigationTakagi2022},
quantum broadcasting
\cite{VQB,unilocal,NoPracticalXiaoyunlong2025}, 
quantum resource distillation \cite{VRD,VRDA}, 
estimating two-point correlation functions \cite{buscemi2013twopoint,twopoint2},
and reversing unknown quantum processes \cite{virtualcombs}. 
Experimentally, they can be implemented by sampling from a set of quantum
operations, i.e., completely positive trace-preserving (CPTP) maps, followed by
postprocessing measurement statistics of the output states 
\cite{HPTPfornonMarkovian,experimentalVRD2024,experimentalVQBzheng2025}.

We will first show that virtual quantum operations indeed enable the perfect
and deterministic cloning of nonorthogonal states, while their linearity
still prohibits perfect universal cloning of quantum states.
Second, we derive a necessary and sufficient condition for a set of quantum
states to be virtually clonable.
Third, by analyzing the simulation costs, we demonstrate that identifying the
optimal virtual cloning can be efficiently formulated as a semidefinite
program (SDP). Finally, we establish a connection between virtual cloning and
state discrimination, which leads to universal bounds on the optimal cloning
cost.

\section{Criterion for virtual cloning}

The no-cloning theorem states that for any two distinct states $\rho_1$ and
$\rho_2$, no quantum operation $\Lambda$ can satisfy
$\Lambda(\rho_i)=\rho_i\otimes\rho_i$ for both $i=1,2$ unless $\rho_1$ and
$\rho_2$ are orthogonal, i.e., $\rho_1\rho_2=0$.
This can be easily seen from the monotonicity of the trace norm
\cite{Nielsen_Chuang_2010}.
Suppose that there exists a quantum operation $\Lambda$ such that
$\Lambda(\rho_i)=\rho_i\otimes\rho_i$;
then applying the operation $\Lambda$ $n-1$ times gives us the $1\to n$ cloning
$\Lambda^{n-1}(\rho_i)=\rho_i^{\otimes n}$.
We denote the trace norm by $\norm{\cdot}$, defined as
$\norm{A}=\Tr(\sqrt{A^\dagger A})$; then the monotonicity of
the trace norm under quantum operations implies that
\begin{equation}
\norm{\rho_1^{\otimes n}-\rho_2^{\otimes n}}
=\norm{\Lambda^{n-1}(\rho_1)-\Lambda^{n-1}(\rho_2)}
\le\norm{\rho_1-\rho_2}.
\end{equation}
The distinguishability of $\rho_1^{\otimes n}$ and $\rho_2^{\otimes n}$ 
when $n\to\infty$ implies that $\lim_{n\to\infty}
\norm{\rho_1^{\otimes n}-\rho_2^{\otimes n}}=2$.
Therefore, $\norm{\rho_1-\rho_2}\ge 2$; i.e.,
$\rho_1$ and $\rho_2$ must be orthogonal.

The complete positivity of quantum operations prohibits the cloning of
nonorthogonal states. Therefore, a natural question is whether quantum cloning
is possible without requiring the complete positivity.
The answer is still negative if we relax the complete positivity to positivity
because the monotonicity of the trace norm still holds for positive
trace-preserving maps \cite{M.B.RUSKAI1994}.
Thus, in this work we consider the cloning under HPTP maps, which we call
virtual cloning.
In recent years, lifting quantum operations characterized by CPTP maps to
virtual operations characterized by HPTP maps has drawn a lot of research
interest.
Theoretically, virtual operations can  accomplish quantum information processing
tasks previously deemed impossible \cite{VQB,VRD,virtualcombs}. Experimentally,
the virtual operations can be physically realized across various quantum systems
\cite{HPTPfornonMarkovian,experimentalVRD2024,experimentalVQBzheng2025}.

To illustrate the advantage of virtual cloning, we demonstrate that it is
possible to clone the nonorthogonal states $\ket{0}$ and $\ket{+}$, a task that
is  impossible in the standard cloning scenario. By definition, we want to find
an HPTP map $\tL$ such that $\tL(\ketbra{0})=\ketbra{0}\otimes\ketbra{0}$ and
$\tL(\ketbra{+})=\ketbra{+}\otimes\ketbra{+}$. One can easily verify that the
linear map determined by $\tL(\I_2)=\frac{1}{2}\I_2\otimes\I_2$ and
$\tL(\sigma_i)=\frac{1}{2}(\sigma_i\otimes\I_2
+\I_2\otimes\sigma_i+\sigma_i\otimes\sigma_i)$
readily fulfills this requirement, where $\I_2$ is the
identity matrix and $\sigma_i\in\{\sigma_x,\sigma_y,\sigma_z\}$
are the Pauli matrices.

This example illustrates that lifting the positivity constraint enables the
cloning of nonorthogonal states, significantly expanding the scope of
quantum cloning. This naturally prompts several key questions: 
Is simultaneous virtual cloning feasible for arbitrary state pairs?
What constitutes the maximal virtually clonable state set?
What are the necessary and sufficient conditions for virtual cloning?
Given that multiple cloning operations exist, which is optimal?

It is important to note that, unlike quantum broadcasting \cite{VQB},
universal cloning remains impossible even with virtual operations.
This is evident from the states $\ketbra{0}$, $\ketbra{1}$, and $\I/2$.
The linearity of $\tL$ implies that
$\tL(\I)=\tL(\ketbra{0})+\tL(\ketbra{1})$.
If universal cloning were possible, then
$\I/2\otimes\I/2=(\ketbra{0}\otimes\ketbra{0}
+\ketbra{1}\otimes\ketbra{1})/2$, a clear contradiction.
This highlights two fundamental aspects of virtual cloning: 
One is that a virtual-cloning operation must be defined within
a specific set of quantum states. 
Thus, we formally define virtual cloning as follows:
$\tL$ is a virtual-cloning operation for a set of quantum states $\stateset$,
if it satisfies 
\begin{equation}
   \tL(\rho_i)=\rho_i\otimes\rho_i 
\end{equation}
for all $\rho_i$. The other is that linear independence plays a crucial role in
determining whether a set of states is virtually clonable. In fact, linear
independence provides a necessary and sufficient criterion for virtual cloning.

\begin{theorem}\label{thm:existence}
For a set of quantum states $\{\rho_1,\rho_2,\dots,\rho_m\}$ in a
$d$-dimensional quantum system, a virtual-cloning operation exists if and only if
the set of the states is linearly independent.
\end{theorem}

\textit{Proof.}
One can easily show that the linear independence is a sufficient
condition for them to be virtually clonable. The linear independence of
$\{\rho_1,\rho_2,\dots,\rho_m\}$ implies that $m\le d^2$, which
is the dimension of the Hermitian operator space. Without loss
of generality, we can always assume $m=d^2$; otherwise, extra states
$\rho_{m+1},\dots,\rho_{d^2}$ can be added to the set such that
the linear independence still holds. Then the linear independence
would imply that the $d^2$ states form a basis; thus, a linear map
$\tL$ would be uniquely identified by $\tL(\rho_i)=\rho_i\otimes\rho_i$
for $i=1,2,\dots,d^2$ and one can easily verify that $\tL$ is HPTP.

To show that the linear independence condition is also necessary, 
we suppose that a virtual-cloning  operation $\tL$ exists for a set of
linearly dependent states $\stateset$. The linear dependence implies that 
$r_i\in\dR$ exist such that 
\begin{equation}\label{eq:dependence}
\sum_{i=1}^m r_i \rho_i =0,
\end{equation}
where not all $r_i$ are zero.
Applying the operation $\tL$ $n-1$ times gives us the $1\to n$ virtual cloning
$\tL^{n-1}(\rho_i)=\rho_i^{\otimes n}$.
Applying the $1\to n$ cloning operation $\tL^{n-1}$ to both sides of 
Eq.~(\ref{eq:dependence}) yields the key equation for deriving the
contradiction:
\begin{equation}
    \sum_{i=1}^m r_i \rho_i^{\otimes n}=0.
    \label{eq:rhon}
\end{equation}
As the set $S=\bigcup_{i \neq j}\{X : \Tr[X(\rho_i - \rho_j)]=0\}$ is a finite
union of proper subspaces, it is always of measure zero. Thus, 
a Hermitian operator $Y$ that is not in $S$ always exists; i.e.,
$\Tr[Y(\rho_i - \rho_j)] \neq 0$ for all $i \neq j$.
This implies that all numbers $y_i = \Tr(Y \rho_i)$ are distinct. 
Furthermore, Eq.~\eqref{eq:rhon} implies that
$\sum_{i=1}^m r_i\Tr(\rho_i^{\otimes n}Y^{\otimes n})=0$, i.e.,
\begin{equation}
\sum_{i=1}^my_i^nr_i=\sum_{i=1}^m M_{ni}r_i=0
\label{eq:vondemonde}
\end{equation}
for $n=0,1,2,\dots,m-1$
\footnote{The case $n=0$ is obtained by taking the trace of
Eq.~\eqref{eq:dependence}.}, 
where the square matrix $M$ defined by $M_{ni}=y_i^n$
for $n=0,1,\dots,m-1$ and $i=1,2,\dots,m$ is a so-called Vandermonde
matrix and it is invertible when all $y_i$ are distinct
\cite{MatrixAnalysisHorn}. Thus, Eq.~\eqref{eq:vondemonde} implies that
all $r_i=0$, which contradicts the linear-dependence assumption.

The above theorem shows the advantages of virtual cloning in three aspects. 
First, unlike most quantum cloning machines that are designed for pure states, 
a virtual-cloning machine can clone any set of pure or mixed states, 
provided it is linearly independent.
Especially, any two distinct states $\rho_1$ and $\rho_2$ can be virtually
cloned simultaneously. 
We also highlight that the linear independence presented in
Theorem~\ref{thm:existence} differs from that in probabilistic cloning
\cite{CloningProbabilisticDuan,CloningProbabilisticPati}:
in the former, it applies to density operators, whereas in the latter,
it applies to pure states (i.e., kets).
Second, the virtual-cloning machine in Theorem \ref{thm:existence} is both
perfect and deterministic for the corresponding set of states. Therefore, the
linear independence of a set  of states also implies the existence of
a $1\to n$ perfect and deterministic virtual-cloning  machine.
Third, according to Theorem~\ref{thm:existence}, for a $d$-dimensional
system, at most $d^2$ quantum states can be virtually cloned simultaneously.
This restriction can be lifted when considering the $k\to n$ cloning.
More precisely, for any finite set of states, a finite
number $k$ always exists such that the virtual $k\to n$ cloning is possible,
where $n$ can be arbitrarily large.
See Appendix~\ref{app:k2ncloning} for more details.

\section{Optimal virtual cloning}

For a given set of quantum states, we have presented a criterion
for the existence of virtual-cloning operations. 
However, when virtual-cloning operations are not unique, 
we want to find a way to choose the optimal one.
To this end, we need to introduce the concept of simulation cost for virtual
operations. We start by clarifying how a virtual operation is implemented in
experiments. Mathematically, a virtual operation $\tL$ always admits a
decomposition \cite{Jiangphysicalimplement}
\begin{equation}
    \tL=\lambda_+ \Lambda_+-\lambda_-\Lambda_-,
    \label{eq:qpd}
\end{equation}
where $\Lambda_\pm$ are CPTP maps, $\lambda_\pm\ge0$ and $\lambda_+-\lambda_-=1$.
For a virtual operation $\tL$, our aim is not to obtain the complete information about
the final state $\tL(\rho)$; instead we are interested in only the partial
information revealed by some observables $X_1,X_2,\dots,X_\ell$, i.e., the
expected values $\Tr[\tL(\rho)X_1],\Tr[\tL(\rho)X_2],\dots,\Tr[\tL(\rho)X_\ell]$. 
Most practical quantum protocols exhibit this characteristic, as information
about their final output is accessible only through measurements.

For simplicity, we consider simulating the measurement of an observable $X$ with
eigenvalues $\pm 1$
\footnote{For the general observable $X$, the results are similar,
but the influence of the factor $\eta$ may be state dependent.}.
The virtual operation $\tL$ can be realized
by decomposing it into two quantum operations $\Lambda_+$ and $\Lambda_-$,
as given by Eq.~(\ref{eq:qpd}).
For each input state $\rho$, we perform the 
quantum operations $\Lambda_{\pm}$ with probabilities
$p_{\pm}=\lambda_{\pm}/\eta$, where $\eta=\lambda_++\lambda_-$.
In the $i$th round, if $\Lambda_+$ is performed,
we multiply the measurement result $x_i$ of
$X$ by a factor $\eta$, i.e., $\hat{x}_i=\eta x_i$;
if $\Lambda_-$ is performed,
we multiply the measurement result $x_i$ of
$X$ by a factor $-\eta$, i.e., $\hat{x}_i=-\eta x_i$.
After $N$ rounds of measurements, the expected value $\Tr[\tL(\rho)X]$
can be estimated from $\frac{1}{N}\sum_{i=1}^N \hat{x}_i$.
Mathematically, $\hat{x}$ is an estimator with
\begin{equation}
P(\hat{x}=\pm\eta)=p_+\Tr[\Lambda_+(\rho)X_\pm]
+p_-\Tr[\Lambda_-(\rho)X_\mp],
\end{equation}
where $X_\pm=(\I\pm X)/2$.
One can easily verify that
\begin{equation}
    \expval*{\hat{x}}
    =\expval*{X},
    \quad
    \expval*{\hat{x}^2}
    =\eta^2\expval*{X^2},
    \label{eq:xhat}
\end{equation}
where $\expval{\cdot}$ denotes the expected value.
Thus, the estimator $\hat{x}$ will give the expectation value of $X$,
but the variance will be larger than that of directly measuring $X$
due to the overhead factor $\eta^2$ in Eq.~\eqref{eq:xhat}.
This means that as the factor $\eta$ increases, more rounds $N$ are
required to achieve a specific precision of $\expval*{X}$.
Alternatively, one can also see the statistical influence
of the factor $\eta$ from Hoeffding's inequality \cite{Hoeffding1963}:
\begin{equation}
    P\qty(\abs{\frac{1}{N}\sum_{i=1}^N\hat{x}_i-\expval*{X}}\ge\varepsilon)
    \le 2\exp(-\frac{\varepsilon^2N}{2\eta^2}).
\end{equation}
Therefore, we will call $\eta$ the simulation
cost for the virtual operation $\tL$
corresponding the decomposition $\tL=\lambda_+ \Lambda_+-\lambda_-\Lambda_-$.

With the above discussion, we naturally want to implement a virtual cloning with
the minimal simulation cost.
We note that optimizing the physical implementation of virtual cloning actually
involves two steps. As the decomposition in Eq.~\eqref{eq:qpd} is not unique, we
need to find the optimal decomposition in Eq.~\eqref{eq:qpd} that minimizes
$\eta=\lambda_++\lambda_-$, which we will call the optimal simulation cost
of $\tL$. Hereafter, we will always use $\eta$ to denote the optimal
simulation cost. In addition, we need to optimize all virtual
operations that can implement the virtual-cloning processes.
Mathematically, this is equivalent to the following optimization problem:
\begin{equation}
    \begin{aligned}
        &\minover[\tL,\,\Lambda_\pm,\,\lambda_\pm] && \lambda_++\lambda_-\\
        &~~~\subto && \tL(\rho_i)=\rho_i\otimes\rho_i\quad
        \text{for } i=1,2,\cdots,m,\\
        &&& \tL=\lambda_+\Lambda_+-\lambda_-\Lambda_-,\\
        &&& \lambda_\pm\ge 0,~\Lambda_\pm\text{ are CPTP.}
        \end{aligned}
\end{equation}
We will call the solution the optimal cloning cost for $\stateset$.

By taking advantage of the Choi-Jamio\l kowski isomorphism
\cite{JAMIOLKOWSKI1972,CHOI1975},
the optimal cloning cost can be recast as an SDP. Similar to the CPTP map,
we can also define the Choi matrix for $\tL$ as
\begin{equation}
    J=\Id\otimes\tL(\ketbra{\Omega}),
\end{equation}
where $\ket{\Omega}=\sum_{i=1}^d\ket{i}_R\ket{i}_S$ is an unnormalized maximally
entangled state in $\cH_R\otimes\cH_S$ and $\cH_R=\cH_S=\dC^d$ and $\Id$ is
the identity map.
Note that here $\tL$ is a map from operators on $\cH_S$ to operators on two
copies $\cH_S$. Thus, $J$ is an operator on $\cH_R\otimes\cH_S\otimes\cH_{S'}$.
In fact, $J$ is just an alternative expression for $\tL$
in the sense that
\begin{equation}
    \tL(\rho)=\Tr_R[(\rho^T\otimes\I_{SS'})J].
\end{equation}
Furthermore, the conditions that $\tL$ is Hermitian preserving and
trace preserving are equivalent to the conditions that $J$ is Hermitian
and $\Tr_{SS'}(J)=\I_R$, respectively.
Also, we can write the decomposition in Eq.~\eqref{eq:qpd}
in terms of Choi matrices. To make the constraints linear, we define
$J_\pm=\lambda_\pm\Id\otimes\Lambda_\pm(\ketbra{\Omega})$.
Then, $J=J_+-J_-$, and the CPTP property of $\Lambda_\pm$
is equivalent to the conditions that $J_\pm\ge0$ and
$\Tr_{SS'}(J_\pm)=\lambda_\pm\I_R$.
Therefore, we obtain the following SDP for the optimal cloning cost.

\begin{observation}
The optimal cloning cost for a set of virtually clonable quantum states
$\{\rho_1,\rho_2,\dots,\rho_m\}$ can be obtained from the following SDP:
\begin{equation}
    \begin{aligned}
        &\minover[J_\pm,\,\lambda_\pm] && \lambda_++\lambda_-\\
        &~\subto && \Tr_R[(\rho_i^T\otimes\I_{SS'})(J_+-J_-)]
        =\rho_i\otimes\rho_i,\\
        &&&\Tr_{SS'}(J_+)=\lambda_+\I_R,~\Tr_{SS'}(J_-)=\lambda_-\I_R,\\
        &&&J_+ \ge 0,~J_- \ge 0,
    \end{aligned}
    \label{eq:costSDP}
\end{equation}
where $J_\pm$ are Hermitian matrices defined on the Hilbert space
$\cH_R\otimes\cH_S\otimes\cH_{S'}$ and $\cH_R=\cH_S=\cH_{S'}=\dC^d$.
\end{observation}

\section{Cloning of nonorthogonal state pairs}

Let us consider the simplest case in which we try to clone a pair of distinct states
$\rho_1$ and $\rho_2$. The no-cloning theorem implies that cloning $\rho_1$ and
$\rho_2$ is impossible unless $\rho_1$ and $\rho_2$ are orthogonal
\cite{no-cloning2YUEN1986,nobroadcasting}. On the
contrary, Theorem~\ref{thm:existence} implies that virtual cloning is always
possible. In this section, we study their optimal cloning cost
and how it is related to the problem of quantum state discrimination.
We will use $\eta(\rho_1,\rho_2)$ to denote the optimal cloning cost
for the state pair $\{\rho_1,\rho_2\}$, with no ambiguity.

We begin with the case in which both states are pure, i.e.,
$\rho_1=\ketbra{\psi_1}$ and $\rho_2=\ketbra{\psi_2}$.
In this case, we can analytically solve the optimization problem
in Eq.~\eqref{eq:costSDP}.

\begin{theorem}\label{thm:optimalcosttwostates}
For any two distinct pure quantum states $\ket{\psi_1}$ and $\ket{\psi_2}$, 
their optimal cloning cost is $\eta(\ket{\psi_1},\ket{\psi_2})
=\sqrt{1+\abs{\braket{\psi_1}{\psi_2}}^2}$. Furthermore, the optimal
virtual-cloning process can be done with randomized unitary operations.
\end{theorem}

The proof of Theorem~\ref{thm:optimalcosttwostates} can be found in Appendix~\ref{app:optimalcloning}.
Actually, we prove a more general result on pure-state conversion:
The optimal simulation cost for any pure-state pair conversion
$\ket{\psi_i}\to\ket{\varphi_i}$ for $i=1,2$ is given by
\begin{equation}
    \eta=\sqrt{\frac{1-\abs{\braket{\varphi_1}{\varphi_2}}^2}
    {1-\abs{\braket{\psi_1}{\psi_2}}^2}}
    =\frac{\norm{\ketbra{\varphi_1}-\ketbra{\varphi_2}}}
    {\norm{\ketbra{\psi_1}-\ketbra{\psi_2}}}
    \label{eq:purestatesconversion}
\end{equation}
when the fidelity of the final states is smaller than that of the initial
states, i.e.,
$\abs{\braket{\varphi_1}{\varphi_2}}<\abs{\braket{\psi_1}{\psi_2}}$.

Equation~\eqref{eq:purestatesconversion} gives us more information
on virtual cloning. First, the optimal cloning cost for $1\to n$ virtual
cloning is given by
\begin{equation}
    \eta_{1\to n}(\ket{\psi_1},\ket{\psi_2})
    =\sqrt{\frac{1-\abs{\braket{\psi_1}{\psi_2}}^{2n}}
    {1-\abs{\braket{\psi_1}{\psi_2}}^2}},
\end{equation}
which is bounded even when $n\to\infty$. Second, the last expression
in Eq.~\eqref{eq:purestatesconversion} indicates that the optimal
simulation seems closely related to the quantum state discrimination
problem because for any pair of states $\rho_1$ and $\rho_2$, their
discrimination probability is determined by $\norm{\rho_1-\rho_2}$
\cite{Helstrom1969}. Below, we demonstrate that these two results are indeed
closely related to the quantum discrimination task and further establish
the following bounds on the optimal cloning cost.

\begin{theorem}\label{thm:costbounds}
For any two distinct quantum states $\rho_1$ and $\rho_2$, 
the optimal $1\to n$ cloning cost is bounded by
\begin{equation}
\frac{\norm{\rho_1^{\otimes n}-\rho_2^{\otimes n}}}
{\norm{\rho_1-\rho_2}}\le
\eta_{1\to n}(\rho_1,\rho_2)\le
\frac{4}{\norm{\rho_1-\rho_2}}-1.
\label{eq:costbound}
\end{equation}
\end{theorem}

We start from the lower bound. To simplify the notation, we consider only the
case that $n=2$, and the generalization to general $n$ is trivial. To get the
lower bound of the optimal cloning cost, we consider the dual problem of
Eq.~\eqref{eq:costSDP} \cite{TQIWatrous2018}:
\begin{equation}
    \begin{aligned}
        &\maxover[Y_i, M_\pm] && \sum_{i=1}^m\Tr[(\rho_i\otimes\rho_i)Y_i]\\
        &~\subto && M_-\otimes\I_{SS'}\le 
        \sum_{i=1}^m\rho_i^T\otimes Y_i\le M_+\otimes\I_{SS'},\\
        &&&\Tr(M_-)=-1,~\Tr(M_+)=1,
    \end{aligned}
    \label{eq:costSDPdual}
\end{equation}
where $Y_i$ are Hermitian matrices defined on Hilbert space $\cH_S\otimes\cH_{S'}$
and $M_\pm$ are Hermitian matrices defined on Hilbert space $\cH_R$.
Moreover, the strong duality holds due to Slater's condition 
\cite{ConvexOptimizationBoyd2004},
i.e., the solution also equals the
optimal cloning cost $\eta$. Therefore, any feasible $Y_i$ and $M_\pm$ give
a lower bound of $\eta$. For the problem of lower bounding $\eta(\rho_1,\rho_2)$,
$m$ in Eq.~\eqref{eq:costSDPdual} equals $2$.
Let $\{P_\pm\}$ and $\{Q_\pm\}$ be the optimal measurements
to distinguish $\{\rho_1,\rho_2\}$ and $\{\rho_1\otimes\rho_1,\rho_2\otimes\rho_2\}$,
respectively, i.e.,
\begin{equation}
    \Tr[(P_+-P_-)(\rho_1-\rho_2)]
    =\norm{\rho_1-\rho_2},
    \label{eq:discrimination}
\end{equation}
and similarly for $Q_\pm$ \cite{Helstrom1969}. Now, we choose
\begin{align}
    &Y_1=-Y_2=\frac{Q_+-Q_-}{\norm{\rho_1-\rho_2}},\\
    M_+=&-M_-=\frac{[(P_+-P_-)(\rho_1-\rho_2)]^T}{\norm{\rho_1-\rho_2}}.
\end{align}
One can easily verify that all the constraints in Eq.~\eqref{eq:costSDPdual}
are satisfied, and the objective function gives the
desired lower bound in Theorem~\ref{thm:costbounds}.
Actually, with a similar argument, one can prove that
\begin{equation} \label{nonequalbound}
\eta_{1\to n}(\rho_1,\rho_2)
\ge\frac{\norm{p_1\rho_1^{\otimes n}-p_2\rho_2^{\otimes n}}}
{\norm{p_1\rho_1-p_2\rho_2}}
\end{equation}
for any probability distribution $(p_1,p_2)$, which corresponds to
the state discrimination with non-equal prior probabilities \cite{Helstrom1969}.
See Appendix~\ref{app:nonequalbound} for more details.

The global upper bound in Eq.~\eqref{eq:costbound} is based on the
simple observation that if a virtual operation $\tD$ exists such
that
\begin{equation}
    \tD(\rho_1)=\ketbra{1}\qc
    \tD(\rho_2)=\ketbra{2},
    \label{eq:flag}
\end{equation}
where $\ket{1}$ and $\ket{2}$ are orthogonal, then
$\eta_{1\to n}(\rho_1,\rho_2)$ is no larger than
the optimal simulation cost of $\tD$. This is because a CPTP map $\Lambda$ always exists such that
$\Lambda(\ketbra{i})=\rho_i^{\otimes n}$, and the optimal simulation cost
for $\tL:=\Lambda\circ\tD$ is no larger than that of $\tD$.
One possible choice of $\tD$ is based on the state discrimination
measurement $\{P_\pm\}$ as in Eq.~\eqref{eq:discrimination}.
We choose $\cD_+(\rho)=\Tr(\rho P_+)\ketbra{1}+\Tr(\rho P_-)\ketbra{2}$
and $\cD_-(\rho)=p_1\ketbra{1}+p_2\ketbra{2}$, where the probability
distribution $(p_1,p_2)\propto (\Tr(\rho_2P_+),\Tr(\rho_1P_-))$.
Note also that $\Tr(\rho_2P_+)+\Tr(\rho_1P_-)=1-\norm{\rho_1-\rho_2}/2$.
A direct calculation shows that
$\tD:=\eta_+\cD_+-\eta_-\cD_-$ satisfies Eq.~\eqref{eq:flag},
where $\eta_+=2/\norm{\rho_1-\rho_2}$ and $\eta_-=\eta_+-1$.
Therefore, the optimal simulation cost for $\tD$ is no larger than
$\eta_++\eta_-=4/\norm{\rho_1-\rho_2}-1$, from which the global upper bound
in Eq.~\eqref{eq:costbound} follows.

Actually, the global upper bound for $1\to n$ not only exists for
state pairs but also exists for any virtually clonable set
$\{\rho_1,\rho_2,\dots,\rho_m\}$. This is because we can also construct
a virtual operation such that $\tD(\rho_i)=\ketbra{i}$ for any $i=1,2,\dots,m$,
and similarly, a virtual $1\to n$ cloning operation $\tL$ can be constructed with
optimal simulation cost no larger than that of $\tD$, which is independent of $n$.
This feature highlights the practical significance of the virtual-cloning
protocol.

\section{Conclusion}

The no-cloning theorem, a cornerstone of quantum information theory, asserts
the impossibility of creating an exact copy of nonorthogonal quantum states.
In this work, we introduced a virtual-cloning protocol that circumvents
this limitation. The protocol employs recently trending virtual quantum
processes, implemented via standard quantum operations and postprocessing
of output state measurement results.
Qualitatively, we established a necessary and sufficient
criterion for the existence of a virtual operation capable of simultaneously
cloning a set of states. Quantitatively, we demonstrated that the optimization of a
virtual-cloning protocol can be formulated as a semidefinite programming
problem, and an analytical solution was determined for any pair of pure states.
Finally, we revealed a connection between virtual cloning and quantum state
discrimination, from which we deduced universal bounds on the optimal cloning cost.

Several promising avenues exist for future research.
First, a deeper exploration of the relationship between virtual cloning and
quantum state discrimination is worth further study, particularly for scenarios involving
more than two states and the statistical comparison of these two approaches.
Second, investigating approximate virtual-cloning machines and quantifying their
advantages over established cloning protocols present an interesting direction.
Third, applying the current methodology to quantum certification
\cite{EisertNRP2020},
especially for estimating nonlinear functions
\cite{ElbenPRL2020,PTmomentsYuPRL2021},
constitutes a productive line of research. 
Last, generalizing our results to quantum gates represents
a nontrivial challenge, given the fundamental distinctions between state
and gate replication 
\cite{GatecloningChiribella2008,GatecloningDur2015,GatecloningChiribella2015}.

\begin{acknowledgments}
  This work was supported by
  the National Natural Science Foundation of China
  (Grants No. 12205170 and No. 12174224)
  and the Shandong Provincial Natural Science Foundation of China
  (Grant No. ZR2022QA084).
\end{acknowledgments}

\appendix

\section{$k\to n$ VIRTUAL CLONING}\label{app:k2ncloning}

For any finite set of states $\stateset$, Theorem~\ref{thm:existence} states
that no virtual-cloning operation exists for them if they are linearly
dependent. This can be lifted by inputting more than one but finitely many
copies of these states. More precisely, for any finite set of linearly
dependent states $\stateset$, a finite number $k$ always exists
such that the states in 
$\{ \rho_1^{\otimes k},\rho_2^{\otimes k}, \dots ,\rho_m^{\otimes k} \} $
are linearly independent, then Theorem~\ref{thm:existence} implies that
$k \to n $ virtual cloning for $\stateset$ is possible. 

Suppose that for some $k$, the set 
$\{\rho_1^{\otimes k},\rho_2^{\otimes k},\dots,\rho_m^{\otimes k}\}$ 
remains linearly dependent, 
i.e., $r_i \in \mathbb{R}$ exist such that
$\sum_{i=1}^m r_i \rho_i^{\otimes k} =0$,
where not all $r_i$ are zero. Taking the trace or partial traces yields
\begin{equation}\label{AppA:linearindet}
    \sum_{i=1}^m r_i \rho_i^{\otimes t} =0
\end{equation}
for $t = 0,1,2, \dots ,k $. Like for the proof of Theorem~\ref{thm:existence},
we can always find a Hermitian operator 
$Y$ such that $\Tr[Y(\rho_i-\rho_j)] \ne 0$ for all $i \ne j$. 
This implies all $y_i=\Tr(Y\rho_i)$ are distinct, then 
Eq.~(\ref{AppA:linearindet}) leads to 
$\sum_{i=1}^m y_i^t r_i = \sum_{i=1}^m M_{ti} r_i = 0$
for $t = 0,1,2, \dots ,k $. For any $k \ge m-1$,
we utilize the first $m$ equations, where the matrix $M$ defined as 
$M_{ti} = y_i^t$ for $t= 0,1,\dots,m-1$ and $i=1,2,\dots,m$ 
is a so-called Vandermonde matrix \cite{MatrixAnalysisHorn}. 
As all $y_i$ are distinct, $M$ is invertible. Thus, all $r_i = 0$, 
which contradicts to the linear-dependence assumption. 
Therefore, for any set of $m$ linearly dependent states, we need
at most $m-1$ copies to make them linearly independent.
Therefore, the $(m-1) \to n$ virtual cloning is always possible,
where $n$ can be arbitrarily large.

\section{OPTIMAL $k\to n$ VIRTUAL CLONING}\label{app:optimalcloning}

In this appendix, we will prove the general result.
\begin{theoremb}
For any two pure quantum states $\ket{\psi_+}$ and $\ket{\psi_-}$, 
their optimal $k \to n$ cloning cost is 
$\eta_{k \to n}(\ket{\psi_+},\ket{\psi_-})
=\sqrt{\frac{1-\abs{\braket{\psi_+}{\psi_-}}^{2n} }
{1-\abs{\braket{\psi_+}{\psi_-}}^{2k}}}$.
Furthermore, the optimal
cloning process can be done with randomized unitary operations.
\label{thm:optimalk2n}
\end{theoremb}

Theorem~\ref{thm:optimalk2n} is a direct corollary of the following lemma,
whose proof also provides the explicit form of the virtual operation.
This lemma may be of independent interest for the study of
general virtual quantum operations.

\begin{lemmab}
Let $\ket{\psi_+}$ and $\ket{\psi_-}$ be two distinct states in Hilbert space
$\cH$, $\ket{\varphi_+}$ and $\ket{\varphi_-}$ be two states in Hilbert space
$\cH'$, and the fidelities are $F=\abs{\braket{\psi_+}{\psi_-}}^2$
and $F'=\abs{\braket{\varphi_+}{\varphi_-}}^2$, respectively. Then, the optimal
virtual operation $\tL$ such that $\tL(\ketbra{\psi_\pm})=\ketbra{\varphi_\pm}$
has the optimal simulation cost
$\eta=\max\qty{1,\sqrt{\frac{1-F'}{1-F}}}$.
That is, when $F\le F'$, a CPTP map $\Lambda$ exists such that
$\Lambda(\ketbra{\psi_\pm})=\ketbra{\varphi_\pm}$;
when $F>F'$, $\eta=\sqrt{\frac{1-F'}{1-F}}$.
\end{lemmab}

\textit{Proof.} For the case $F\le F'$, the proof is obvious.
We just need to consider the reduced map of
$U\ket{\psi_\pm}\ket{0}=\ket{\varphi_\pm}\ket{\chi_\pm}$,
where $\ket{\chi_\pm}$ satisfy
$\braket{\chi_+}{\chi_-}=\frac{\braket{\psi_+}{\psi_-}}
{\braket{\varphi_+}{\varphi_-}}$.
Hence, we will mainly focus on the case that $F>F'$.

We start by reforming these four states.
As the transformation $\Lambda(\ketbra{\psi_\pm})=\ketbra{\varphi_\pm}$
does not depend on the global phases, we can always choose suitable phases
such that
$\braket{\psi_+}{\psi_-}\ge 0$ and $\braket{\varphi_+}{\varphi_-}\ge 0$.
Thus, we can parametrize $\ket{\psi_\pm}$ as
$\alpha\ket{\psi_0}\pm\beta\ket{\psi_1}$, with $\alpha\ge\beta\ge0$,
and $\{\ket{\psi_0},\ket{\psi_1}\}$
is a suitable basis. Similarly, we can parametrize $\ket{\varphi_\pm}$ as
$\ket{\varphi_\pm}= a\ket{\varphi_0}\pm b\ket{\varphi_1}$, with $a\ge b\ge0$. 
On the one hand, for any virtual operation $\tL$ 
satisfying $ \tL(\ketbra{\psi_\pm})=\ketbra{\varphi_\pm}$, 
we can always construct another virtual operation
\begin{equation}\label{eqa:tLprime}
    \tL'= \Lambda' \circ  \tL \circ \Lambda,
\end{equation}
where $\Lambda$ and $\Lambda'$ are CPTP and map operators on $\dC^2$ to
operators on $\cH$ and operators on $\cH'$ to operators on $\dC^2$,
respectively. Furthermore,
$\Lambda(\rho) = K\rho K^\dagger$ and
$\Lambda'(\rho) = K'\rho K'^{\dagger} + \Tr(\rho P'_{\perp})\ketbra{0}$,
where $K=\ketbra{\psi_0}{0}+\ketbra{\psi_1}{1}$,
$K'=\ketbra{0}{\varphi_0}+\ketbra{1}{\varphi_1}$,
and $P'_{\perp}=\I_{\cH'}-\ketbra{\varphi_0}-\ketbra{\varphi_1}$.
The resulting $\tL'$ in Eq.~\eqref{eqa:tLprime}
maps states $\alpha\ket{0}\pm\beta\ket{1}$ in $\dC^2$ to 
states $a\ket{0}\pm b\ket{1}$ in $\dC^2$,
and the optimal simulation cost of $\tL'$ is no larger than that of $\tL$.
On the other hand, any $\tL'$ that
maps states $\alpha\ket{0}\pm\beta\ket{1}$ in $\dC^2$ to 
states $a\ket{0}\pm b\ket{1}$ in $\dC^2$,
can be extended to a map $\tL$ by
\begin{equation}
    \tL = \cE' \circ  \tL' \circ \cE,
\end{equation}
where
$\cE(\rho)=K^\dagger\rho K + \Tr(\rho P_{\perp})\ketbra{0}$, 
$\cE'(\rho)=K'^{\dagger}\rho K'$,
$P_{\perp}=\I_{\cH}-\ketbra{\psi_0}-\ketbra{\psi_1}$.
The optimal simulation cost of $\tL$ is no larger than that of $\tL'$
because $\cE$ and $\cE'$ are CPTP. Thus, without loss of generality,
we can consider the special case in which $\cH=\cH'=\dC^2$ and
\begin{equation}
    \ket{\psi_\pm}=\alpha\ket{0}\pm\beta\ket{1}\qc
    \ket{\varphi_\pm}=a\ket{0}\pm b\ket{1},
    \label{eqa:states}
\end{equation}
where $\alpha\ge\beta\ge0$ and $a\ge b\ge 0$.
Moreover, $F=1-4\alpha^2\beta^2$ and $F'=1-4a^2b^2$.

The optimal simulation cost $\eta$ is given by
\begin{equation}
    \begin{aligned}
        &\minover[\tL,\,\Lambda_\pm,\,\lambda_\pm] && \lambda_++\lambda_-\\
        &~~~\subto && \tL(\ketbra{\psi_\pm})=\ketbra{\varphi_\pm}\\
        &&& \tL=\lambda_+\Lambda_+-\lambda_-\Lambda_-,\\
        &&& \lambda_\pm\ge 0,~\Lambda_\pm\text{ are CPTP.}
        \end{aligned}
        \label{eqa:opt}
\end{equation}
We will show that the solution of Eq.~\eqref{eqa:opt} equals the solution of
\begin{equation}
    \begin{aligned}
        &\minover[J] && \frac{1}{2}\norm{J}\\
        &~\subto &&
        \Tr_1[(\ketbra{\psi_\pm}\otimes\I)J]=\ketbra{\varphi_\pm},\\
        &&& \Tr_2(J)=\I,
        \end{aligned}
        \label{eqa:relaxJ}
\end{equation}
where $J$ is the corresponding Choi matrix of $\tL$
and $\Tr_1$ and $\Tr_2$ denote the partial traces on the first and second
subsystems, respectively. Note that in the general formula for the relation
between $J$ and $\tL$, one needs to take the transposition for states
$\ketbra{\psi_\pm}$, but under our assumption in Eq.~\eqref{eqa:states},
$\ketbra{\psi_\pm}$ are real symmetric matrices.

As the solution of Eq.~\eqref{eqa:opt} is greater than or equal to
the solution of Eq.~\eqref{eqa:relaxJ} \cite{Jiangphysicalimplement},
we need to show only that equality can be attained
by giving an explicit solution of Eq.~\eqref{eqa:relaxJ}.
The optimization in Eq.~\eqref{eqa:relaxJ} is a convex optimization, thus
we can take advantage of the symmetry to simplify the form of the $4\times 4$
matrix $J$. One can easily see that the objective function and feasible region
are invariant under $J\to J^T$ or
$J\to (\sigma_z\otimes \sigma_z)J(\sigma_z\otimes \sigma_z)$.
Thus, we can assume that $J$ is invariant under these two transformations.
Taking the last constraint of Eq.~\eqref{eqa:relaxJ} into consideration,
we have the following parametrization of $J$
\begin{equation}
    J=\frac{1}{2}\big(\I\otimes\I + r\I\otimes\sigma_z
    +x\sigma_x\otimes\sigma_x
    +y\sigma_y\otimes\sigma_y
    +z\sigma_z\otimes\sigma_z\big),
\end{equation}
where $r,x,y,$ and $z$ are real numbers. Furthermore, the first constraint
in Eq.~\eqref{eqa:relaxJ} implies that
$x=\sqrt{\frac{1-F'}{1-F}}$ and
$z=\sqrt{\frac{F'}{F}}-\frac{r}{\sqrt{F}}$.
To simplify the notation, we define the parameters
$\xi=\sqrt{\frac{1-F'}{1-F}}$, 
$\mu=\sqrt{\frac{F'}{F}}$,
and $\gamma=\frac{r}{\sqrt{F}}$.
One can easily see that $\xi$ and $\mu$ are constants and satisfy
$\xi>1$ and $\mu<1$. Thus, $J$ can be written as
\begin{widetext}
\begin{equation}
J=\frac{1}{2} 
\begin{bmatrix}
    1+\mu-(1-\sqrt{F})\gamma & 0 & 0 & \xi -y \\
 0 & 1-\mu+(1-\sqrt{F})\gamma & \xi +y & 0 \\
 0 & \xi +y & 1-\mu+(1+\sqrt{F})\gamma & 0 \\
 \xi -y & 0 & 0 & 1+\mu-(1+\sqrt{F})\gamma \\
\end{bmatrix},
\end{equation}
\end{widetext}
where $\gamma$ and $y$ are parameters to be optimized.
One can easily verify that for any $2\times 2$ Hermitian matrix
\begin{equation}
    \norm{\mqty[a_{11}&&a_{12}\\a_{21}&&a_{22}]}
    \ge 2 \abs{a_{12}},
\end{equation}
and the equality holds if and only if $a_{11}=a_{22}$ and
$\abs{a_{11}}\le\abs{a_{12}}$. Thus, we get that
$\norm{J}\ge\abs{\xi+y}+\abs{\xi-y}\ge 2\xi$, i.e.,
the solution of Eq.~\eqref{eqa:relaxJ} is no smaller than
$\xi=\sqrt{\frac{1-F'}{1-F}}$.

Finally, we show that a $\tL$ exists such that $\lambda_++\lambda_-=\xi$.
Actually, any HPTP map induced by the Choi matrix $J$ satisfying
\begin{equation}
    \gamma=0\qc
    1-\xi-\mu\le y\le\xi-1-\mu
    \label{eqa:ximucons}
\end{equation}
meets this requirement. Specifically, we take
$\gamma=0$ and $y=1-\xi-\mu$, then the Choi matrix $J$ reads
$J=\frac{\xi+\mu}{2}\ketbra{\Omega}
+\frac{1-\mu}{2}\ketbra{\Omega_x}
-\frac{\xi-1}{2}\ketbra{\Omega_z}$,
where $\ket{\Omega_x}=\ket{01}+\ket{10}$
and $\ket{\Omega_z}=\ket{00}-\ket{11}$.
Thus, the corresponding $\tL$ takes the form
\begin{equation}
    \tL(\rho)
    =\frac{\xi+\mu}{2}\rho
    +\frac{1-\mu}{2}\sigma_x\rho\sigma_x
    -\frac{\xi-1}{2}\sigma_z\rho\sigma_z,
\end{equation}
for which $\lambda_+=\frac{\xi+\mu}{2}+\frac{1-\mu}{2}=\frac{\xi+1}{2}$
and $\lambda_-=\frac{\xi-1}{2}$.
Therefore, we prove that $\lambda_++\lambda_-=\xi=\sqrt{\frac{1-F'}{1-F}}$.

\section{PROOF OF EQUATION~(\ref{nonequalbound})}\label{app:nonequalbound}

As in the main text, we consider only the case that $n=2$, 
and the generalization to general $n$ is trivial.
Let $\{P_\pm\}$ and $\{Q_\pm\}$ be the optimal measurements 
to distinguish 
$\{\rho_1,\rho_2\}$ with prior probabilities $\{p_1,p_2\}$ 
and $\{\rho_1^{\otimes 2},\rho_2^{\otimes 2}\}$ with 
the same prior probabilities $\{p_1,p_2\}$, respectively. 
Now, we let $Y_1=p_1 Y$ and $Y_2=-p_2Y$, where 
\begin{equation}
    Y=\frac{Q_+-Q_-}{\norm{p_1\rho_1-p_2\rho_2}},
\end{equation} 
and we let $M_+ = -M_- = M$, where 
\begin{equation}\label{AppC:M}
    M=\frac{[(P_+-P_-)(p_1\rho_1-p_2\rho_2)]^T}{\norm{p_1\rho_1-p_2\rho_2}}.
\end{equation}
From Helstrom's bound \cite{Helstrom1969}, we have
\begin{align}
    &\Tr[(P_+-P_-)(p_1\rho_1-p_2\rho_2)]
    =\norm{p_1\rho_1-p_2\rho_2},
    \label{AppC:trace1}\\
    &\Tr[(Q_+-Q_-)(p_1\rho_1^{\otimes 2}-p_2\rho_2^{\otimes 2})]
    =\norm{p_1\rho_1^{\otimes 2}-p_2\rho_2^{\otimes 2}}.
    \label{AppC:trace2}
\end{align}
Equation~(\ref{AppC:M}) implies that $M_+ \ge 0$ and $M_- \le 0$,
which further imply that
\begin{equation}
    M_-\otimes\I_{SS'}\le 
        \frac{(p_1\rho_1-p_2\rho_2)^T}{\norm{p_1\rho_1-p_2\rho_2}}
        \otimes(Q_+-Q_-)
        \le M_+\otimes\I_{SS'},
\end{equation}
and Eq.~(\ref{AppC:trace1}) implies that $\Tr(M_{\pm})=\pm1 $ .
Therefore, $M_+,M_-,Y_1,$ and $Y_2$ satisfy all
the constraints in Eq.~\eqref{eq:costSDPdual}. 
Furthermore, from Eq.~(\ref{AppC:trace2}), the objective function
equals
\begin{equation}
    \Tr(\rho_1^{\otimes 2}Y_1+\rho_2^{\otimes 2}Y_2)
    =\frac{\norm{p_1\rho_1^{\otimes 2}-p_2\rho_2^{\otimes 2}}}
    {\norm{p_1\rho_1-p_2\rho_2}}.
\end{equation}
Hence, we complete the proof.

\bibliography{referencearticle}

\begin{thebibliography}{56}%
\makeatletter
\providecommand \@ifxundefined [1]{%
 \@ifx{#1\undefined}
}%
\providecommand \@ifnum [1]{%
 \ifnum #1\expandafter \@firstoftwo
 \else \expandafter \@secondoftwo
 \fi
}%
\providecommand \@ifx [1]{%
 \ifx #1\expandafter \@firstoftwo
 \else \expandafter \@secondoftwo
 \fi
}%
\providecommand \natexlab [1]{#1}%
\providecommand \enquote  [1]{``#1''}%
\providecommand \bibnamefont  [1]{#1}%
\providecommand \bibfnamefont [1]{#1}%
\providecommand \citenamefont [1]{#1}%
\providecommand \href@noop [0]{\@secondoftwo}%
\providecommand \href [0]{\begingroup \@sanitize@url \@href}%
\providecommand \@href[1]{\@@startlink{#1}\@@href}%
\providecommand \@@href[1]{\endgroup#1\@@endlink}%
\providecommand \@sanitize@url [0]{\catcode `\\12\catcode `\$12\catcode
  `\&12\catcode `\#12\catcode `\^12\catcode `\_12\catcode `\%12\relax}%
\providecommand \@@startlink[1]{}%
\providecommand \@@endlink[0]{}%
\providecommand \url  [0]{\begingroup\@sanitize@url \@url }%
\providecommand \@url [1]{\endgroup\@href {#1}{\urlprefix }}%
\providecommand \urlprefix  [0]{URL }%
\providecommand \Eprint [0]{\href }%
\providecommand \doibase [0]{https://doi.org/}%
\providecommand \selectlanguage [0]{\@gobble}%
\providecommand \bibinfo  [0]{\@secondoftwo}%
\providecommand \bibfield  [0]{\@secondoftwo}%
\providecommand \translation [1]{[#1]}%
\providecommand \BibitemOpen [0]{}%
\providecommand \bibitemStop [0]{}%
\providecommand \bibitemNoStop [0]{.\EOS\space}%
\providecommand \EOS [0]{\spacefactor3000\relax}%
\providecommand \BibitemShut  [1]{\csname bibitem#1\endcsname}%
\let\auto@bib@innerbib\@empty
\bibitem [{\citenamefont {Wootters}\ and\ \citenamefont
  {Zurek}(1982)}]{nocloningWootters1982}%
  \BibitemOpen
  \bibfield  {author} {\bibinfo {author} {\bibfnamefont {W.~K.}\ \bibnamefont
  {Wootters}}\ and\ \bibinfo {author} {\bibfnamefont {W.~H.}\ \bibnamefont
  {Zurek}},\ }\bibfield  {title} {\bibinfo {title} {A single quantum cannot be
  cloned},\ }\href {https://doi.org/10.1038/299802a0} {\bibfield  {journal}
  {\bibinfo  {journal} {Nature}\ }\textbf {\bibinfo {volume} {299}},\ \bibinfo
  {pages} {802} (\bibinfo {year} {1982})}\BibitemShut {NoStop}%
\bibitem [{\citenamefont {Dieks}(1982)}]{nosuperluminalDIEKS1982}%
  \BibitemOpen
  \bibfield  {author} {\bibinfo {author} {\bibfnamefont {D.}~\bibnamefont
  {Dieks}},\ }\bibfield  {title} {\bibinfo {title} {Communication by {EPR}
  devices},\ }\href
  {https://doi.org/https://doi.org/10.1016/0375-9601(82)90084-6} {\bibfield
  {journal} {\bibinfo  {journal} {Phys. Lett. A}\ }\textbf {\bibinfo {volume}
  {92}},\ \bibinfo {pages} {271} (\bibinfo {year} {1982})}\BibitemShut
  {NoStop}%
\bibitem [{\citenamefont {Scarani}\ \emph {et~al.}(2005)\citenamefont
  {Scarani}, \citenamefont {Iblisdir}, \citenamefont {Gisin},\ and\
  \citenamefont {Ac\'{\i}n}}]{CloningRevModPhys2005}%
  \BibitemOpen
  \bibfield  {author} {\bibinfo {author} {\bibfnamefont {V.}~\bibnamefont
  {Scarani}}, \bibinfo {author} {\bibfnamefont {S.}~\bibnamefont {Iblisdir}},
  \bibinfo {author} {\bibfnamefont {N.}~\bibnamefont {Gisin}},\ and\ \bibinfo
  {author} {\bibfnamefont {A.}~\bibnamefont {Ac\'{\i}n}},\ }\bibfield  {title}
  {\bibinfo {title} {Quantum cloning},\ }\href
  {https://doi.org/10.1103/RevModPhys.77.1225} {\bibfield  {journal} {\bibinfo
  {journal} {Rev. Mod. Phys.}\ }\textbf {\bibinfo {volume} {77}},\ \bibinfo
  {pages} {1225} (\bibinfo {year} {2005})}\BibitemShut {NoStop}%
\bibitem [{\citenamefont {Fan}\ \emph {et~al.}(2014)\citenamefont {Fan},
  \citenamefont {Wang}, \citenamefont {Jing}, \citenamefont {Yue},
  \citenamefont {Shi}, \citenamefont {Zhang},\ and\ \citenamefont
  {Mu}}]{CloningreviewFAN2015}%
  \BibitemOpen
  \bibfield  {author} {\bibinfo {author} {\bibfnamefont {H.}~\bibnamefont
  {Fan}}, \bibinfo {author} {\bibfnamefont {Y.-N.}\ \bibnamefont {Wang}},
  \bibinfo {author} {\bibfnamefont {L.}~\bibnamefont {Jing}}, \bibinfo {author}
  {\bibfnamefont {J.-D.}\ \bibnamefont {Yue}}, \bibinfo {author} {\bibfnamefont
  {H.-D.}\ \bibnamefont {Shi}}, \bibinfo {author} {\bibfnamefont {Y.-L.}\
  \bibnamefont {Zhang}},\ and\ \bibinfo {author} {\bibfnamefont {L.-Z.}\
  \bibnamefont {Mu}},\ }\bibfield  {title} {\bibinfo {title} {Quantum cloning
  machines and the applications},\ }\href
  {https://doi.org/https://doi.org/10.1016/j.physrep.2014.06.004} {\bibfield
  {journal} {\bibinfo  {journal} {Phys. Rep.}\ }\textbf {\bibinfo {volume}
  {544}},\ \bibinfo {pages} {241} (\bibinfo {year} {2014})}\BibitemShut
  {NoStop}%
\bibitem [{\citenamefont {Bu\ifmmode~\check{z}\else \v{z}\fi{}ek}\ and\
  \citenamefont {Hillery}(1996)}]{CloningOptimal1-2UQCM}%
  \BibitemOpen
  \bibfield  {author} {\bibinfo {author} {\bibfnamefont {V.}~\bibnamefont
  {Bu\ifmmode~\check{z}\else \v{z}\fi{}ek}}\ and\ \bibinfo {author}
  {\bibfnamefont {M.}~\bibnamefont {Hillery}},\ }\bibfield  {title} {\bibinfo
  {title} {Quantum copying: Beyond the no-cloning theorem},\ }\href
  {https://doi.org/10.1103/PhysRevA.54.1844} {\bibfield  {journal} {\bibinfo
  {journal} {Phys. Rev. A}\ }\textbf {\bibinfo {volume} {54}},\ \bibinfo
  {pages} {1844} (\bibinfo {year} {1996})}\BibitemShut {NoStop}%
\bibitem [{\citenamefont {Gisin}\ and\ \citenamefont
  {Massar}(1997)}]{CloningOptimalN-MUQCM}%
  \BibitemOpen
  \bibfield  {author} {\bibinfo {author} {\bibfnamefont {N.}~\bibnamefont
  {Gisin}}\ and\ \bibinfo {author} {\bibfnamefont {S.}~\bibnamefont {Massar}},\
  }\bibfield  {title} {\bibinfo {title} {Optimal quantum cloning machines},\
  }\href {https://doi.org/10.1103/PhysRevLett.79.2153} {\bibfield  {journal}
  {\bibinfo  {journal} {Phys. Rev. Lett.}\ }\textbf {\bibinfo {volume} {79}},\
  \bibinfo {pages} {2153} (\bibinfo {year} {1997})}\BibitemShut {NoStop}%
\bibitem [{\citenamefont {Bu\ifmmode~\check{z}\else \v{z}\fi{}ek}\ and\
  \citenamefont {Hillery}(1998)}]{Cloning1-2UQCMd-dim}%
  \BibitemOpen
  \bibfield  {author} {\bibinfo {author} {\bibfnamefont {V.}~\bibnamefont
  {Bu\ifmmode~\check{z}\else \v{z}\fi{}ek}}\ and\ \bibinfo {author}
  {\bibfnamefont {M.}~\bibnamefont {Hillery}},\ }\bibfield  {title} {\bibinfo
  {title} {Universal optimal cloning of arbitrary quantum states: From qubits
  to quantum registers},\ }\href {https://doi.org/10.1103/PhysRevLett.81.5003}
  {\bibfield  {journal} {\bibinfo  {journal} {Phys. Rev. Lett.}\ }\textbf
  {\bibinfo {volume} {81}},\ \bibinfo {pages} {5003} (\bibinfo {year}
  {1998})}\BibitemShut {NoStop}%
\bibitem [{\citenamefont {Niu}\ and\ \citenamefont
  {Griffiths}(1998)}]{CloningAsymmetricNiu1998}%
  \BibitemOpen
  \bibfield  {author} {\bibinfo {author} {\bibfnamefont {C.-S.}\ \bibnamefont
  {Niu}}\ and\ \bibinfo {author} {\bibfnamefont {R.~B.}\ \bibnamefont
  {Griffiths}},\ }\bibfield  {title} {\bibinfo {title} {Optimal copying of one
  quantum bit},\ }\href {https://doi.org/10.1103/PhysRevA.58.4377} {\bibfield
  {journal} {\bibinfo  {journal} {Phys. Rev. A}\ }\textbf {\bibinfo {volume}
  {58}},\ \bibinfo {pages} {4377} (\bibinfo {year} {1998})}\BibitemShut
  {NoStop}%
\bibitem [{\citenamefont {Cerf}(2000)}]{CloningAsymmetricPauli2000}%
  \BibitemOpen
  \bibfield  {author} {\bibinfo {author} {\bibfnamefont {N.~J.}\ \bibnamefont
  {Cerf}},\ }\bibfield  {title} {\bibinfo {title} {Pauli cloning of a quantum
  bit},\ }\href {https://doi.org/10.1103/PhysRevLett.84.4497} {\bibfield
  {journal} {\bibinfo  {journal} {Phys. Rev. Lett.}\ }\textbf {\bibinfo
  {volume} {84}},\ \bibinfo {pages} {4497} (\bibinfo {year}
  {2000})}\BibitemShut {NoStop}%
\bibitem [{\citenamefont {Duan}\ and\ \citenamefont
  {Guo}(1998)}]{CloningProbabilisticDuan}%
  \BibitemOpen
  \bibfield  {author} {\bibinfo {author} {\bibfnamefont {L.-M.}\ \bibnamefont
  {Duan}}\ and\ \bibinfo {author} {\bibfnamefont {G.-C.}\ \bibnamefont {Guo}},\
  }\bibfield  {title} {\bibinfo {title} {Probabilistic cloning and
  identification of linearly independent quantum states},\ }\href
  {https://doi.org/10.1103/PhysRevLett.80.4999} {\bibfield  {journal} {\bibinfo
   {journal} {Phys. Rev. Lett.}\ }\textbf {\bibinfo {volume} {80}},\ \bibinfo
  {pages} {4999} (\bibinfo {year} {1998})}\BibitemShut {NoStop}%
\bibitem [{\citenamefont {Pati}(1999)}]{CloningProbabilisticPati}%
  \BibitemOpen
  \bibfield  {author} {\bibinfo {author} {\bibfnamefont {A.~K.}\ \bibnamefont
  {Pati}},\ }\bibfield  {title} {\bibinfo {title} {Quantum superposition of
  multiple clones and the novel cloning machine},\ }\href
  {https://doi.org/10.1103/PhysRevLett.83.2849} {\bibfield  {journal} {\bibinfo
   {journal} {Phys. Rev. Lett.}\ }\textbf {\bibinfo {volume} {83}},\ \bibinfo
  {pages} {2849} (\bibinfo {year} {1999})}\BibitemShut {NoStop}%
\bibitem [{\citenamefont {Bru\ss{}}\ \emph {et~al.}(2000)\citenamefont
  {Bru\ss{}}, \citenamefont {Cinchetti}, \citenamefont {Mauro~D'Ariano},\ and\
  \citenamefont {Macchiavello}}]{CloningPhaseBru2000}%
  \BibitemOpen
  \bibfield  {author} {\bibinfo {author} {\bibfnamefont {D.}~\bibnamefont
  {Bru\ss{}}}, \bibinfo {author} {\bibfnamefont {M.}~\bibnamefont {Cinchetti}},
  \bibinfo {author} {\bibfnamefont {G.}~\bibnamefont {Mauro~D'Ariano}},\ and\
  \bibinfo {author} {\bibfnamefont {C.}~\bibnamefont {Macchiavello}},\
  }\bibfield  {title} {\bibinfo {title} {Phase-covariant quantum cloning},\
  }\href {https://doi.org/10.1103/PhysRevA.62.012302} {\bibfield  {journal}
  {\bibinfo  {journal} {Phys. Rev. A}\ }\textbf {\bibinfo {volume} {62}},\
  \bibinfo {pages} {012302} (\bibinfo {year} {2000})}\BibitemShut {NoStop}%
\bibitem [{\citenamefont {Fan}\ \emph {et~al.}(2001)\citenamefont {Fan},
  \citenamefont {Matsumoto}, \citenamefont {Wang},\ and\ \citenamefont
  {Wadati}}]{CloningPhaseFan2001}%
  \BibitemOpen
  \bibfield  {author} {\bibinfo {author} {\bibfnamefont {H.}~\bibnamefont
  {Fan}}, \bibinfo {author} {\bibfnamefont {K.}~\bibnamefont {Matsumoto}},
  \bibinfo {author} {\bibfnamefont {X.-B.}\ \bibnamefont {Wang}},\ and\
  \bibinfo {author} {\bibfnamefont {M.}~\bibnamefont {Wadati}},\ }\bibfield
  {title} {\bibinfo {title} {Quantum cloning machines for equatorial qubits},\
  }\href {https://doi.org/10.1103/PhysRevA.65.012304} {\bibfield  {journal}
  {\bibinfo  {journal} {Phys. Rev. A}\ }\textbf {\bibinfo {volume} {65}},\
  \bibinfo {pages} {012304} (\bibinfo {year} {2001})}\BibitemShut {NoStop}%
\bibitem [{\citenamefont {Niu}\ and\ \citenamefont
  {Griffiths}(1999)}]{CloningPhaseNiu}%
  \BibitemOpen
  \bibfield  {author} {\bibinfo {author} {\bibfnamefont {C.-S.}\ \bibnamefont
  {Niu}}\ and\ \bibinfo {author} {\bibfnamefont {R.~B.}\ \bibnamefont
  {Griffiths}},\ }\bibfield  {title} {\bibinfo {title} {Two-qubit copying
  machine for economical quantum eavesdropping},\ }\href
  {https://doi.org/10.1103/PhysRevA.60.2764} {\bibfield  {journal} {\bibinfo
  {journal} {Phys. Rev. A}\ }\textbf {\bibinfo {volume} {60}},\ \bibinfo
  {pages} {2764} (\bibinfo {year} {1999})}\BibitemShut {NoStop}%
\bibitem [{\citenamefont {Bru\ss{}}\ and\ \citenamefont
  {Macchiavello}(2002)}]{EavesdroppingBruss2002}%
  \BibitemOpen
  \bibfield  {author} {\bibinfo {author} {\bibfnamefont {D.}~\bibnamefont
  {Bru\ss{}}}\ and\ \bibinfo {author} {\bibfnamefont {C.}~\bibnamefont
  {Macchiavello}},\ }\bibfield  {title} {\bibinfo {title} {Optimal
  eavesdropping in cryptography with three-dimensional quantum states},\ }\href
  {https://doi.org/10.1103/PhysRevLett.88.127901} {\bibfield  {journal}
  {\bibinfo  {journal} {Phys. Rev. Lett.}\ }\textbf {\bibinfo {volume} {88}},\
  \bibinfo {pages} {127901} (\bibinfo {year} {2002})}\BibitemShut {NoStop}%
\bibitem [{\citenamefont {Cerf}\ \emph {et~al.}(2002)\citenamefont {Cerf},
  \citenamefont {Bourennane}, \citenamefont {Karlsson},\ and\ \citenamefont
  {Gisin}}]{EavesdroppingCerf2002}%
  \BibitemOpen
  \bibfield  {author} {\bibinfo {author} {\bibfnamefont {N.~J.}\ \bibnamefont
  {Cerf}}, \bibinfo {author} {\bibfnamefont {M.}~\bibnamefont {Bourennane}},
  \bibinfo {author} {\bibfnamefont {A.}~\bibnamefont {Karlsson}},\ and\
  \bibinfo {author} {\bibfnamefont {N.}~\bibnamefont {Gisin}},\ }\bibfield
  {title} {\bibinfo {title} {Security of quantum key distribution using
  $\mathit{d}$-level systems},\ }\href
  {https://doi.org/10.1103/PhysRevLett.88.127902} {\bibfield  {journal}
  {\bibinfo  {journal} {Phys. Rev. Lett.}\ }\textbf {\bibinfo {volume} {88}},\
  \bibinfo {pages} {127902} (\bibinfo {year} {2002})}\BibitemShut {NoStop}%
\bibitem [{\citenamefont {Bru\ss{}}\ \emph {et~al.}(1998)\citenamefont
  {Bru\ss{}}, \citenamefont {Ekert},\ and\ \citenamefont
  {Macchiavello}}]{StateEstiBru1998}%
  \BibitemOpen
  \bibfield  {author} {\bibinfo {author} {\bibfnamefont {D.}~\bibnamefont
  {Bru\ss{}}}, \bibinfo {author} {\bibfnamefont {A.}~\bibnamefont {Ekert}},\
  and\ \bibinfo {author} {\bibfnamefont {C.}~\bibnamefont {Macchiavello}},\
  }\bibfield  {title} {\bibinfo {title} {Optimal universal quantum cloning and
  state estimation},\ }\href {https://doi.org/10.1103/PhysRevLett.81.2598}
  {\bibfield  {journal} {\bibinfo  {journal} {Phys. Rev. Lett.}\ }\textbf
  {\bibinfo {volume} {81}},\ \bibinfo {pages} {2598} (\bibinfo {year}
  {1998})}\BibitemShut {NoStop}%
\bibitem [{\citenamefont {Brougham}\ \emph {et~al.}(2006)\citenamefont
  {Brougham}, \citenamefont {Andersson},\ and\ \citenamefont
  {Barnett}}]{MeasuCompaBrougham2006}%
  \BibitemOpen
  \bibfield  {author} {\bibinfo {author} {\bibfnamefont {T.}~\bibnamefont
  {Brougham}}, \bibinfo {author} {\bibfnamefont {E.}~\bibnamefont
  {Andersson}},\ and\ \bibinfo {author} {\bibfnamefont {S.~M.}\ \bibnamefont
  {Barnett}},\ }\bibfield  {title} {\bibinfo {title} {Cloning and joint
  measurements of incompatible components of spin},\ }\href
  {https://doi.org/10.1103/PhysRevA.73.062319} {\bibfield  {journal} {\bibinfo
  {journal} {Phys. Rev. A}\ }\textbf {\bibinfo {volume} {73}},\ \bibinfo
  {pages} {062319} (\bibinfo {year} {2006})}\BibitemShut {NoStop}%
\bibitem [{\citenamefont {Sagawa}\ and\ \citenamefont
  {Ueda}(2008)}]{MeasuCompaSagawa2008}%
  \BibitemOpen
  \bibfield  {author} {\bibinfo {author} {\bibfnamefont {T.}~\bibnamefont
  {Sagawa}}\ and\ \bibinfo {author} {\bibfnamefont {M.}~\bibnamefont {Ueda}},\
  }\bibfield  {title} {\bibinfo {title} {Accuracy matrix in a generalized
  simultaneous measurement of a qubit system},\ }\href
  {https://doi.org/10.1103/PhysRevA.77.012313} {\bibfield  {journal} {\bibinfo
  {journal} {Phys. Rev. A}\ }\textbf {\bibinfo {volume} {77}},\ \bibinfo
  {pages} {012313} (\bibinfo {year} {2008})}\BibitemShut {NoStop}%
\bibitem [{\citenamefont {Thekkadath}\ \emph {et~al.}(2017)\citenamefont
  {Thekkadath}, \citenamefont {Saaltink}, \citenamefont {Giner},\ and\
  \citenamefont {Lundeen}}]{MeasuCompaThekk2017}%
  \BibitemOpen
  \bibfield  {author} {\bibinfo {author} {\bibfnamefont {G.~S.}\ \bibnamefont
  {Thekkadath}}, \bibinfo {author} {\bibfnamefont {R.~Y.}\ \bibnamefont
  {Saaltink}}, \bibinfo {author} {\bibfnamefont {L.}~\bibnamefont {Giner}},\
  and\ \bibinfo {author} {\bibfnamefont {J.~S.}\ \bibnamefont {Lundeen}},\
  }\bibfield  {title} {\bibinfo {title} {Determining complementary properties
  with quantum clones},\ }\href
  {https://doi.org/10.1103/PhysRevLett.119.050405} {\bibfield  {journal}
  {\bibinfo  {journal} {Phys. Rev. Lett.}\ }\textbf {\bibinfo {volume} {119}},\
  \bibinfo {pages} {050405} (\bibinfo {year} {2017})}\BibitemShut {NoStop}%
\bibitem [{\citenamefont {Bruß}\ and\ \citenamefont
  {Macchiavello}(2003)}]{Cloning-EntanglementBruss2003}%
  \BibitemOpen
  \bibfield  {author} {\bibinfo {author} {\bibfnamefont {D.}~\bibnamefont
  {Bruß}}\ and\ \bibinfo {author} {\bibfnamefont {C.}~\bibnamefont
  {Macchiavello}},\ }\bibfield  {title} {\bibinfo {title} {On the entanglement
  structure in quantum cloning},\ }\href
  {https://doi.org/10.1023/A:1026061117305} {\bibfield  {journal} {\bibinfo
  {journal} {Found. Phys.}\ }\textbf {\bibinfo {volume} {33}},\ \bibinfo
  {pages} {1617} (\bibinfo {year} {2003})}\BibitemShut {NoStop}%
\bibitem [{\citenamefont {Lamoureux}\ \emph {et~al.}(2004)\citenamefont
  {Lamoureux}, \citenamefont {Navez}, \citenamefont
  {Fiur\'a\ifmmode~\check{s}\else \v{s}\fi{}ek},\ and\ \citenamefont
  {Cerf}}]{Cloning-EntanglementLamoureux2004}%
  \BibitemOpen
  \bibfield  {author} {\bibinfo {author} {\bibfnamefont {L.-P.}\ \bibnamefont
  {Lamoureux}}, \bibinfo {author} {\bibfnamefont {P.}~\bibnamefont {Navez}},
  \bibinfo {author} {\bibfnamefont {J.}~\bibnamefont
  {Fiur\'a\ifmmode~\check{s}\else \v{s}\fi{}ek}},\ and\ \bibinfo {author}
  {\bibfnamefont {N.~J.}\ \bibnamefont {Cerf}},\ }\bibfield  {title} {\bibinfo
  {title} {Cloning the entanglement of a pair of quantum bits},\ }\href
  {https://doi.org/10.1103/PhysRevA.69.040301} {\bibfield  {journal} {\bibinfo
  {journal} {Phys. Rev. A}\ }\textbf {\bibinfo {volume} {69}},\ \bibinfo
  {pages} {040301} (\bibinfo {year} {2004})}\BibitemShut {NoStop}%
\bibitem [{\citenamefont {Weedbrook}\ \emph {et~al.}(2008)\citenamefont
  {Weedbrook}, \citenamefont {Grosse}, \citenamefont {Symul}, \citenamefont
  {Lam},\ and\ \citenamefont {Ralph}}]{Cloning-EntanglementWeedbrook2008}%
  \BibitemOpen
  \bibfield  {author} {\bibinfo {author} {\bibfnamefont {C.}~\bibnamefont
  {Weedbrook}}, \bibinfo {author} {\bibfnamefont {N.~B.}\ \bibnamefont
  {Grosse}}, \bibinfo {author} {\bibfnamefont {T.}~\bibnamefont {Symul}},
  \bibinfo {author} {\bibfnamefont {P.~K.}\ \bibnamefont {Lam}},\ and\ \bibinfo
  {author} {\bibfnamefont {T.~C.}\ \bibnamefont {Ralph}},\ }\bibfield  {title}
  {\bibinfo {title} {Quantum cloning of continuous-variable entangled states},\
  }\href {https://doi.org/10.1103/PhysRevA.77.052313} {\bibfield  {journal}
  {\bibinfo  {journal} {Phys. Rev. A}\ }\textbf {\bibinfo {volume} {77}},\
  \bibinfo {pages} {052313} (\bibinfo {year} {2008})}\BibitemShut {NoStop}%
\bibitem [{\citenamefont {Yuen}(1986)}]{no-cloning2YUEN1986}%
  \BibitemOpen
  \bibfield  {author} {\bibinfo {author} {\bibfnamefont {H.~P.}\ \bibnamefont
  {Yuen}},\ }\bibfield  {title} {\bibinfo {title} {Amplification of quantum
  states and noiseless photon amplifiers},\ }\href
  {https://doi.org/https://doi.org/10.1016/0375-9601(86)90660-2} {\bibfield
  {journal} {\bibinfo  {journal} {Phys. Lett. A}\ }\textbf {\bibinfo {volume}
  {113}},\ \bibinfo {pages} {405} (\bibinfo {year} {1986})}\BibitemShut
  {NoStop}%
\bibitem [{\citenamefont {Barnum}\ \emph {et~al.}(1996)\citenamefont {Barnum},
  \citenamefont {Caves}, \citenamefont {Fuchs}, \citenamefont {Jozsa},\ and\
  \citenamefont {Schumacher}}]{nobroadcasting}%
  \BibitemOpen
  \bibfield  {author} {\bibinfo {author} {\bibfnamefont {H.}~\bibnamefont
  {Barnum}}, \bibinfo {author} {\bibfnamefont {C.~M.}\ \bibnamefont {Caves}},
  \bibinfo {author} {\bibfnamefont {C.~A.}\ \bibnamefont {Fuchs}}, \bibinfo
  {author} {\bibfnamefont {R.}~\bibnamefont {Jozsa}},\ and\ \bibinfo {author}
  {\bibfnamefont {B.}~\bibnamefont {Schumacher}},\ }\bibfield  {title}
  {\bibinfo {title} {Noncommuting mixed states cannot be broadcast},\ }\href
  {https://doi.org/10.1103/PhysRevLett.76.2818} {\bibfield  {journal} {\bibinfo
   {journal} {Phys. Rev. Lett.}\ }\textbf {\bibinfo {volume} {76}},\ \bibinfo
  {pages} {2818} (\bibinfo {year} {1996})}\BibitemShut {NoStop}%
\bibitem [{\citenamefont {Temme}\ \emph {et~al.}(2017)\citenamefont {Temme},
  \citenamefont {Bravyi},\ and\ \citenamefont
  {Gambetta}}]{ErrorMitigationShortDepthCircuits}%
  \BibitemOpen
  \bibfield  {author} {\bibinfo {author} {\bibfnamefont {K.}~\bibnamefont
  {Temme}}, \bibinfo {author} {\bibfnamefont {S.}~\bibnamefont {Bravyi}},\ and\
  \bibinfo {author} {\bibfnamefont {J.~M.}\ \bibnamefont {Gambetta}},\
  }\bibfield  {title} {\bibinfo {title} {Error mitigation for short-depth
  quantum circuits},\ }\href {https://doi.org/10.1103/PhysRevLett.119.180509}
  {\bibfield  {journal} {\bibinfo  {journal} {Phys. Rev. Lett.}\ }\textbf
  {\bibinfo {volume} {119}},\ \bibinfo {pages} {180509} (\bibinfo {year}
  {2017})}\BibitemShut {NoStop}%
\bibitem [{\citenamefont {Jiang}\ \emph {et~al.}(2021)\citenamefont {Jiang},
  \citenamefont {Wang},\ and\ \citenamefont {Wang}}]{Jiangphysicalimplement}%
  \BibitemOpen
  \bibfield  {author} {\bibinfo {author} {\bibfnamefont {J.}~\bibnamefont
  {Jiang}}, \bibinfo {author} {\bibfnamefont {K.}~\bibnamefont {Wang}},\ and\
  \bibinfo {author} {\bibfnamefont {X.}~\bibnamefont {Wang}},\ }\bibfield
  {title} {\bibinfo {title} {Physical implementability of linear maps and its
  application in error mitigation},\ }\href
  {https://doi.org/10.22331/q-2021-12-07-600} {\bibfield  {journal} {\bibinfo
  {journal} {{Quantum}}\ }\textbf {\bibinfo {volume} {5}},\ \bibinfo {pages}
  {600} (\bibinfo {year} {2021})}\BibitemShut {NoStop}%
\bibitem [{\citenamefont {Takagi}\ \emph {et~al.}(2022)\citenamefont {Takagi},
  \citenamefont {Endo}, \citenamefont {Minagawa},\ and\ \citenamefont
  {Gu}}]{ErrorMitigationTakagi2022}%
  \BibitemOpen
  \bibfield  {author} {\bibinfo {author} {\bibfnamefont {R.}~\bibnamefont
  {Takagi}}, \bibinfo {author} {\bibfnamefont {S.}~\bibnamefont {Endo}},
  \bibinfo {author} {\bibfnamefont {S.}~\bibnamefont {Minagawa}},\ and\
  \bibinfo {author} {\bibfnamefont {M.}~\bibnamefont {Gu}},\ }\bibfield
  {title} {\bibinfo {title} {Fundamental limits of quantum error mitigation},\
  }\href {https://doi.org/10.1038/s41534-022-00618-z} {\bibfield  {journal}
  {\bibinfo  {journal} {npj Quantum Inf.}\ }\textbf {\bibinfo {volume} {8}},\
  \bibinfo {pages} {114} (\bibinfo {year} {2022})}\BibitemShut {NoStop}%
\bibitem [{\citenamefont {Parzygnat}\ \emph {et~al.}(2024)\citenamefont
  {Parzygnat}, \citenamefont {Fullwood}, \citenamefont {Buscemi},\ and\
  \citenamefont {Chiribella}}]{VQB}%
  \BibitemOpen
  \bibfield  {author} {\bibinfo {author} {\bibfnamefont {A.~J.}\ \bibnamefont
  {Parzygnat}}, \bibinfo {author} {\bibfnamefont {J.}~\bibnamefont {Fullwood}},
  \bibinfo {author} {\bibfnamefont {F.}~\bibnamefont {Buscemi}},\ and\ \bibinfo
  {author} {\bibfnamefont {G.}~\bibnamefont {Chiribella}},\ }\bibfield  {title}
  {\bibinfo {title} {Virtual quantum broadcasting},\ }\href
  {https://doi.org/10.1103/PhysRevLett.132.110203} {\bibfield  {journal}
  {\bibinfo  {journal} {Phys. Rev. Lett.}\ }\textbf {\bibinfo {volume} {132}},\
  \bibinfo {pages} {110203} (\bibinfo {year} {2024})}\BibitemShut {NoStop}%
\bibitem [{\citenamefont {Yao}\ \emph {et~al.}(2024)\citenamefont {Yao},
  \citenamefont {Liu}, \citenamefont {Zhu},\ and\ \citenamefont
  {Wang}}]{unilocal}%
  \BibitemOpen
  \bibfield  {author} {\bibinfo {author} {\bibfnamefont {H.}~\bibnamefont
  {Yao}}, \bibinfo {author} {\bibfnamefont {X.}~\bibnamefont {Liu}}, \bibinfo
  {author} {\bibfnamefont {C.}~\bibnamefont {Zhu}},\ and\ \bibinfo {author}
  {\bibfnamefont {X.}~\bibnamefont {Wang}},\ }\bibfield  {title} {\bibinfo
  {title} {Optimal unilocal virtual quantum broadcasting},\ }\href
  {https://doi.org/10.1103/PhysRevA.110.012458} {\bibfield  {journal} {\bibinfo
   {journal} {Phys. Rev. A}\ }\textbf {\bibinfo {volume} {110}},\ \bibinfo
  {pages} {012458} (\bibinfo {year} {2024})}\BibitemShut {NoStop}%
\bibitem [{\citenamefont {Xiao}\ \emph {et~al.}()\citenamefont {Xiao},
  \citenamefont {Liu},\ and\ \citenamefont {Liu}}]{NoPracticalXiaoyunlong2025}%
  \BibitemOpen
  \bibfield  {author} {\bibinfo {author} {\bibfnamefont {Y.}~\bibnamefont
  {Xiao}}, \bibinfo {author} {\bibfnamefont {X.}~\bibnamefont {Liu}},\ and\
  \bibinfo {author} {\bibfnamefont {Z.}~\bibnamefont {Liu}},\ }\href@noop {}
  {\bibinfo {title} {No practical quantum broadcasting: Even virtually}},\
  \Eprint {https://arxiv.org/abs/2503.16380} {arXiv:2503.16380} \BibitemShut
  {NoStop}%
\bibitem [{\citenamefont {Yuan}\ \emph {et~al.}(2024)\citenamefont {Yuan},
  \citenamefont {Regula}, \citenamefont {Takagi},\ and\ \citenamefont
  {Gu}}]{VRD}%
  \BibitemOpen
  \bibfield  {author} {\bibinfo {author} {\bibfnamefont {X.}~\bibnamefont
  {Yuan}}, \bibinfo {author} {\bibfnamefont {B.}~\bibnamefont {Regula}},
  \bibinfo {author} {\bibfnamefont {R.}~\bibnamefont {Takagi}},\ and\ \bibinfo
  {author} {\bibfnamefont {M.}~\bibnamefont {Gu}},\ }\bibfield  {title}
  {\bibinfo {title} {Virtual quantum resource distillation},\ }\href
  {https://doi.org/10.1103/PhysRevLett.132.050203} {\bibfield  {journal}
  {\bibinfo  {journal} {Phys. Rev. Lett.}\ }\textbf {\bibinfo {volume} {132}},\
  \bibinfo {pages} {050203} (\bibinfo {year} {2024})}\BibitemShut {NoStop}%
\bibitem [{\citenamefont {Takagi}\ \emph {et~al.}(2024)\citenamefont {Takagi},
  \citenamefont {Yuan}, \citenamefont {Regula},\ and\ \citenamefont
  {Gu}}]{VRDA}%
  \BibitemOpen
  \bibfield  {author} {\bibinfo {author} {\bibfnamefont {R.}~\bibnamefont
  {Takagi}}, \bibinfo {author} {\bibfnamefont {X.}~\bibnamefont {Yuan}},
  \bibinfo {author} {\bibfnamefont {B.}~\bibnamefont {Regula}},\ and\ \bibinfo
  {author} {\bibfnamefont {M.}~\bibnamefont {Gu}},\ }\bibfield  {title}
  {\bibinfo {title} {Virtual quantum resource distillation: General framework
  and applications},\ }\href {https://doi.org/10.1103/PhysRevA.109.022403}
  {\bibfield  {journal} {\bibinfo  {journal} {Phys. Rev. A}\ }\textbf {\bibinfo
  {volume} {109}},\ \bibinfo {pages} {022403} (\bibinfo {year}
  {2024})}\BibitemShut {NoStop}%
\bibitem [{\citenamefont {Buscemi}\ \emph {et~al.}()\citenamefont {Buscemi},
  \citenamefont {Dall'Arno}, \citenamefont {Ozawa},\ and\ \citenamefont
  {Vedral}}]{buscemi2013twopoint}%
  \BibitemOpen
  \bibfield  {author} {\bibinfo {author} {\bibfnamefont {F.}~\bibnamefont
  {Buscemi}}, \bibinfo {author} {\bibfnamefont {M.}~\bibnamefont {Dall'Arno}},
  \bibinfo {author} {\bibfnamefont {M.}~\bibnamefont {Ozawa}},\ and\ \bibinfo
  {author} {\bibfnamefont {V.}~\bibnamefont {Vedral}},\ }\href@noop {}
  {\bibinfo {title} {Direct observation of any two-point quantum correlation
  function}},\ \Eprint {https://arxiv.org/abs/1312.4240} {arXiv:1312.4240}
  \BibitemShut {NoStop}%
\bibitem [{\citenamefont {Buscemi}\ \emph {et~al.}(2014)\citenamefont
  {Buscemi}, \citenamefont {Dall'Arno}, \citenamefont {Ozawa},\ and\
  \citenamefont {Vedral}}]{twopoint2}%
  \BibitemOpen
  \bibfield  {author} {\bibinfo {author} {\bibfnamefont {F.}~\bibnamefont
  {Buscemi}}, \bibinfo {author} {\bibfnamefont {M.}~\bibnamefont {Dall'Arno}},
  \bibinfo {author} {\bibfnamefont {M.}~\bibnamefont {Ozawa}},\ and\ \bibinfo
  {author} {\bibfnamefont {V.}~\bibnamefont {Vedral}},\ }\bibfield  {title}
  {\bibinfo {title} {Universal optimal quantum correlator},\ }\href
  {https://doi.org/10.1142/S0219749915600023} {\bibfield  {journal} {\bibinfo
  {journal} {Int. J. Quantum Inf.}\ }\textbf {\bibinfo {volume} {12}},\
  \bibinfo {pages} {1560002} (\bibinfo {year} {2014})}\BibitemShut {NoStop}%
\bibitem [{\citenamefont {Zhu}\ \emph {et~al.}(2024)\citenamefont {Zhu},
  \citenamefont {Mo}, \citenamefont {Chen},\ and\ \citenamefont
  {Wang}}]{virtualcombs}%
  \BibitemOpen
  \bibfield  {author} {\bibinfo {author} {\bibfnamefont {C.}~\bibnamefont
  {Zhu}}, \bibinfo {author} {\bibfnamefont {Y.}~\bibnamefont {Mo}}, \bibinfo
  {author} {\bibfnamefont {Y.-A.}\ \bibnamefont {Chen}},\ and\ \bibinfo
  {author} {\bibfnamefont {X.}~\bibnamefont {Wang}},\ }\bibfield  {title}
  {\bibinfo {title} {Reversing unknown quantum processes via virtual combs for
  channels with limited information},\ }\href
  {https://doi.org/10.1103/PhysRevLett.133.030801} {\bibfield  {journal}
  {\bibinfo  {journal} {Phys. Rev. Lett.}\ }\textbf {\bibinfo {volume} {133}},\
  \bibinfo {pages} {030801} (\bibinfo {year} {2024})}\BibitemShut {NoStop}%
\bibitem [{\citenamefont {Rossini}\ \emph {et~al.}(2023)\citenamefont
  {Rossini}, \citenamefont {Maile}, \citenamefont {Ankerhold},\ and\
  \citenamefont {Donvil}}]{HPTPfornonMarkovian}%
  \BibitemOpen
  \bibfield  {author} {\bibinfo {author} {\bibfnamefont {M.}~\bibnamefont
  {Rossini}}, \bibinfo {author} {\bibfnamefont {D.}~\bibnamefont {Maile}},
  \bibinfo {author} {\bibfnamefont {J.}~\bibnamefont {Ankerhold}},\ and\
  \bibinfo {author} {\bibfnamefont {B.~I.~C.}\ \bibnamefont {Donvil}},\
  }\bibfield  {title} {\bibinfo {title} {Single-qubit error mitigation by
  simulating non-{M}arkovian dynamics},\ }\href
  {https://doi.org/10.1103/PhysRevLett.131.110603} {\bibfield  {journal}
  {\bibinfo  {journal} {Phys. Rev. Lett.}\ }\textbf {\bibinfo {volume} {131}},\
  \bibinfo {pages} {110603} (\bibinfo {year} {2023})}\BibitemShut {NoStop}%
\bibitem [{\citenamefont {Zhang}\ \emph {et~al.}(2024)\citenamefont {Zhang},
  \citenamefont {Zhang}, \citenamefont {Liu}, \citenamefont {Fang},
  \citenamefont {Zhang}, \citenamefont {Yuan},\ and\ \citenamefont
  {Lu}}]{experimentalVRD2024}%
  \BibitemOpen
  \bibfield  {author} {\bibinfo {author} {\bibfnamefont {T.}~\bibnamefont
  {Zhang}}, \bibinfo {author} {\bibfnamefont {Y.}~\bibnamefont {Zhang}},
  \bibinfo {author} {\bibfnamefont {L.}~\bibnamefont {Liu}}, \bibinfo {author}
  {\bibfnamefont {X.-X.}\ \bibnamefont {Fang}}, \bibinfo {author}
  {\bibfnamefont {Q.-X.}\ \bibnamefont {Zhang}}, \bibinfo {author}
  {\bibfnamefont {X.}~\bibnamefont {Yuan}},\ and\ \bibinfo {author}
  {\bibfnamefont {H.}~\bibnamefont {Lu}},\ }\bibfield  {title} {\bibinfo
  {title} {Experimental virtual distillation of entanglement and coherence},\
  }\href {https://doi.org/10.1103/PhysRevLett.132.180201} {\bibfield  {journal}
  {\bibinfo  {journal} {Phys. Rev. Lett.}\ }\textbf {\bibinfo {volume} {132}},\
  \bibinfo {pages} {180201} (\bibinfo {year} {2024})}\BibitemShut {NoStop}%
\bibitem [{\citenamefont {Zheng}\ \emph {et~al.}()\citenamefont {Zheng},
  \citenamefont {Nie}, \citenamefont {Liu}, \citenamefont {Luo}, \citenamefont
  {Lu},\ and\ \citenamefont {Liu}}]{experimentalVQBzheng2025}%
  \BibitemOpen
  \bibfield  {author} {\bibinfo {author} {\bibfnamefont {Y.}~\bibnamefont
  {Zheng}}, \bibinfo {author} {\bibfnamefont {X.}~\bibnamefont {Nie}}, \bibinfo
  {author} {\bibfnamefont {H.}~\bibnamefont {Liu}}, \bibinfo {author}
  {\bibfnamefont {Y.}~\bibnamefont {Luo}}, \bibinfo {author} {\bibfnamefont
  {D.}~\bibnamefont {Lu}},\ and\ \bibinfo {author} {\bibfnamefont
  {X.}~\bibnamefont {Liu}},\ }\href@noop {} {\bibinfo {title} {Experimental
  virtual quantum broadcasting}},\ \Eprint {https://arxiv.org/abs/2501.11390}
  {arXiv:2501.11390} \BibitemShut {NoStop}%
\bibitem [{\citenamefont {Nielsen}\ and\ \citenamefont
  {Chuang}(2010)}]{Nielsen_Chuang_2010}%
  \BibitemOpen
  \bibfield  {author} {\bibinfo {author} {\bibfnamefont {M.~A.}\ \bibnamefont
  {Nielsen}}\ and\ \bibinfo {author} {\bibfnamefont {I.~L.}\ \bibnamefont
  {Chuang}},\ }\href@noop {} {\emph {\bibinfo {title} {Quantum Computation and
  Quantum Information, 10th anniversary ed.}}}\ (\bibinfo  {publisher}
  {Cambridge University Press},\ \bibinfo {address} {Cambridge},\ \bibinfo
  {year} {2010})\BibitemShut {NoStop}%
\bibitem [{\citenamefont {Ruskai}(1994)}]{M.B.RUSKAI1994}%
  \BibitemOpen
  \bibfield  {author} {\bibinfo {author} {\bibfnamefont {M.~B.}\ \bibnamefont
  {Ruskai}},\ }\bibfield  {title} {\bibinfo {title} {Beyond strong
  subadditivity? {I}mproved bounds on the contraction of generalized relative
  entropy},\ }\href {https://doi.org/10.1142/S0129055X94000407} {\bibfield
  {journal} {\bibinfo  {journal} {Rev. Math. Phys.}\ }\textbf {\bibinfo
  {volume} {06}},\ \bibinfo {pages} {1147} (\bibinfo {year}
  {1994})}\BibitemShut {NoStop}%
\bibitem [{Note1()}]{Note1}%
  \BibitemOpen
  \bibinfo {note} {The case $n=0$ is obtained by taking the trace of
  Eq.~\protect \eqref {eq:dependence}.}\BibitemShut {Stop}%
\bibitem [{\citenamefont {Horn}\ and\ \citenamefont
  {Johnson}(1985)}]{MatrixAnalysisHorn}%
  \BibitemOpen
  \bibfield  {author} {\bibinfo {author} {\bibfnamefont {R.~A.}\ \bibnamefont
  {Horn}}\ and\ \bibinfo {author} {\bibfnamefont {C.~R.}\ \bibnamefont
  {Johnson}},\ }\href@noop {} {\emph {\bibinfo {title} {Matrix Analysis}}}\
  (\bibinfo  {publisher} {Cambridge University Press},\ \bibinfo {address}
  {Cambridge},\ \bibinfo {year} {1985})\BibitemShut {NoStop}%
\bibitem [{Note2()}]{Note2}%
  \BibitemOpen
  \bibinfo {note} {For the general observable $X$, the results are similar, but
  the influence of the factor $\eta $ may be state dependent.}\BibitemShut
  {Stop}%
\bibitem [{\citenamefont {Hoeffding}(1963)}]{Hoeffding1963}%
  \BibitemOpen
  \bibfield  {author} {\bibinfo {author} {\bibfnamefont {W.}~\bibnamefont
  {Hoeffding}},\ }\bibfield  {title} {\bibinfo {title} {Probability
  inequalities for sums of bounded random variables},\ }\href
  {https://doi.org/10.1080/01621459.1963.10500830} {\bibfield  {journal}
  {\bibinfo  {journal} {J. Am. Stat. Assoc.}\ }\textbf {\bibinfo {volume}
  {58}},\ \bibinfo {pages} {13} (\bibinfo {year} {1963})}\BibitemShut {NoStop}%
\bibitem [{\citenamefont {Jamiołkowski}(1972)}]{JAMIOLKOWSKI1972}%
  \BibitemOpen
  \bibfield  {author} {\bibinfo {author} {\bibfnamefont {A.}~\bibnamefont
  {Jamiołkowski}},\ }\bibfield  {title} {\bibinfo {title} {Linear
  transformations which preserve trace and positive semidefiniteness of
  operators},\ }\href
  {https://doi.org/https://doi.org/10.1016/0034-4877(72)90011-0} {\bibfield
  {journal} {\bibinfo  {journal} {Rep. Math. Phys.}\ }\textbf {\bibinfo
  {volume} {3}},\ \bibinfo {pages} {275} (\bibinfo {year} {1972})}\BibitemShut
  {NoStop}%
\bibitem [{\citenamefont {Choi}(1975)}]{CHOI1975}%
  \BibitemOpen
  \bibfield  {author} {\bibinfo {author} {\bibfnamefont {M.-D.}\ \bibnamefont
  {Choi}},\ }\bibfield  {title} {\bibinfo {title} {Completely positive linear
  maps on complex matrices},\ }\href
  {https://doi.org/https://doi.org/10.1016/0024-3795(75)90075-0} {\bibfield
  {journal} {\bibinfo  {journal} {Linear Algebra Appl}\ }\textbf {\bibinfo
  {volume} {10}},\ \bibinfo {pages} {285} (\bibinfo {year} {1975})}\BibitemShut
  {NoStop}%
\bibitem [{\citenamefont {Helstrom}(1969)}]{Helstrom1969}%
  \BibitemOpen
  \bibfield  {author} {\bibinfo {author} {\bibfnamefont {C.~W.}\ \bibnamefont
  {Helstrom}},\ }\bibfield  {title} {\bibinfo {title} {Quantum detection and
  estimation theory},\ }\href {https://doi.org/10.1007/BF01007479} {\bibfield
  {journal} {\bibinfo  {journal} {J. Stat. Phys.}\ }\textbf {\bibinfo {volume}
  {1}},\ \bibinfo {pages} {231} (\bibinfo {year} {1969})}\BibitemShut {NoStop}%
\bibitem [{\citenamefont {Watrous}(2018)}]{TQIWatrous2018}%
  \BibitemOpen
  \bibfield  {author} {\bibinfo {author} {\bibfnamefont {J.}~\bibnamefont
  {Watrous}},\ }\href@noop {} {\emph {\bibinfo {title} {The Theory of Quantum
  Information}}}\ (\bibinfo  {publisher} {Cambridge University Press},\
  \bibinfo {address} {Cambridge},\ \bibinfo {year} {2018})\BibitemShut
  {NoStop}%
\bibitem [{\citenamefont {Boyd}\ and\ \citenamefont
  {Vandenberghe}(2004)}]{ConvexOptimizationBoyd2004}%
  \BibitemOpen
  \bibfield  {author} {\bibinfo {author} {\bibfnamefont {S.}~\bibnamefont
  {Boyd}}\ and\ \bibinfo {author} {\bibfnamefont {L.}~\bibnamefont
  {Vandenberghe}},\ }\href@noop {} {\emph {\bibinfo {title} {Convex
  Optimization}}}\ (\bibinfo  {publisher} {Cambridge University Press},\
  \bibinfo {address} {Cambridge},\ \bibinfo {year} {2004})\BibitemShut
  {NoStop}%
\bibitem [{\citenamefont {Eisert}\ \emph {et~al.}(2020)\citenamefont {Eisert},
  \citenamefont {Hangleiter}, \citenamefont {Walk}, \citenamefont {Roth},
  \citenamefont {Markham}, \citenamefont {Parekh}, \citenamefont {Chabaud},\
  and\ \citenamefont {Kashefi}}]{EisertNRP2020}%
  \BibitemOpen
  \bibfield  {author} {\bibinfo {author} {\bibfnamefont {J.}~\bibnamefont
  {Eisert}}, \bibinfo {author} {\bibfnamefont {D.}~\bibnamefont {Hangleiter}},
  \bibinfo {author} {\bibfnamefont {N.}~\bibnamefont {Walk}}, \bibinfo {author}
  {\bibfnamefont {I.}~\bibnamefont {Roth}}, \bibinfo {author} {\bibfnamefont
  {D.}~\bibnamefont {Markham}}, \bibinfo {author} {\bibfnamefont
  {R.}~\bibnamefont {Parekh}}, \bibinfo {author} {\bibfnamefont
  {U.}~\bibnamefont {Chabaud}},\ and\ \bibinfo {author} {\bibfnamefont
  {E.}~\bibnamefont {Kashefi}},\ }\bibfield  {title} {\bibinfo {title} {Quantum
  certification and benchmarking},\ }\href
  {https://doi.org/10.1038/s42254-020-0186-4} {\bibfield  {journal} {\bibinfo
  {journal} {Nat. Rev. Phys.}\ }\textbf {\bibinfo {volume} {2}},\ \bibinfo
  {pages} {382} (\bibinfo {year} {2020})}\BibitemShut {NoStop}%
\bibitem [{\citenamefont {Elben}\ \emph {et~al.}(2020)\citenamefont {Elben},
  \citenamefont {Kueng}, \citenamefont {Huang}, \citenamefont {van Bijnen},
  \citenamefont {Kokail}, \citenamefont {Dalmonte}, \citenamefont {Calabrese},
  \citenamefont {Kraus}, \citenamefont {Preskill}, \citenamefont {Zoller},\
  and\ \citenamefont {Vermersch}}]{ElbenPRL2020}%
  \BibitemOpen
  \bibfield  {author} {\bibinfo {author} {\bibfnamefont {A.}~\bibnamefont
  {Elben}}, \bibinfo {author} {\bibfnamefont {R.}~\bibnamefont {Kueng}},
  \bibinfo {author} {\bibfnamefont {H.-Y.~R.}\ \bibnamefont {Huang}}, \bibinfo
  {author} {\bibfnamefont {R.}~\bibnamefont {van Bijnen}}, \bibinfo {author}
  {\bibfnamefont {C.}~\bibnamefont {Kokail}}, \bibinfo {author} {\bibfnamefont
  {M.}~\bibnamefont {Dalmonte}}, \bibinfo {author} {\bibfnamefont
  {P.}~\bibnamefont {Calabrese}}, \bibinfo {author} {\bibfnamefont
  {B.}~\bibnamefont {Kraus}}, \bibinfo {author} {\bibfnamefont
  {J.}~\bibnamefont {Preskill}}, \bibinfo {author} {\bibfnamefont
  {P.}~\bibnamefont {Zoller}},\ and\ \bibinfo {author} {\bibfnamefont
  {B.}~\bibnamefont {Vermersch}},\ }\bibfield  {title} {\bibinfo {title}
  {Mixed-state entanglement from local randomized measurements},\ }\href
  {https://doi.org/10.1103/PhysRevLett.125.200501} {\bibfield  {journal}
  {\bibinfo  {journal} {Phys. Rev. Lett.}\ }\textbf {\bibinfo {volume} {125}},\
  \bibinfo {pages} {200501} (\bibinfo {year} {2020})}\BibitemShut {NoStop}%
\bibitem [{\citenamefont {Yu}\ \emph {et~al.}(2021)\citenamefont {Yu},
  \citenamefont {Imai},\ and\ \citenamefont {G\"uhne}}]{PTmomentsYuPRL2021}%
  \BibitemOpen
  \bibfield  {author} {\bibinfo {author} {\bibfnamefont {X.-D.}\ \bibnamefont
  {Yu}}, \bibinfo {author} {\bibfnamefont {S.}~\bibnamefont {Imai}},\ and\
  \bibinfo {author} {\bibfnamefont {O.}~\bibnamefont {G\"uhne}},\ }\bibfield
  {title} {\bibinfo {title} {Optimal entanglement certification from moments of
  the partial transpose},\ }\href
  {https://doi.org/10.1103/PhysRevLett.127.060504} {\bibfield  {journal}
  {\bibinfo  {journal} {Phys. Rev. Lett.}\ }\textbf {\bibinfo {volume} {127}},\
  \bibinfo {pages} {060504} (\bibinfo {year} {2021})}\BibitemShut {NoStop}%
\bibitem [{\citenamefont {Chiribella}\ \emph {et~al.}(2008)\citenamefont
  {Chiribella}, \citenamefont {D'Ariano},\ and\ \citenamefont
  {Perinotti}}]{GatecloningChiribella2008}%
  \BibitemOpen
  \bibfield  {author} {\bibinfo {author} {\bibfnamefont {G.}~\bibnamefont
  {Chiribella}}, \bibinfo {author} {\bibfnamefont {G.~M.}\ \bibnamefont
  {D'Ariano}},\ and\ \bibinfo {author} {\bibfnamefont {P.}~\bibnamefont
  {Perinotti}},\ }\bibfield  {title} {\bibinfo {title} {Optimal cloning of
  unitary transformation},\ }\href
  {https://doi.org/10.1103/PhysRevLett.101.180504} {\bibfield  {journal}
  {\bibinfo  {journal} {Phys. Rev. Lett.}\ }\textbf {\bibinfo {volume} {101}},\
  \bibinfo {pages} {180504} (\bibinfo {year} {2008})}\BibitemShut {NoStop}%
\bibitem [{\citenamefont {D\"ur}\ \emph {et~al.}(2015)\citenamefont {D\"ur},
  \citenamefont {Sekatski},\ and\ \citenamefont
  {Skotiniotis}}]{GatecloningDur2015}%
  \BibitemOpen
  \bibfield  {author} {\bibinfo {author} {\bibfnamefont {W.}~\bibnamefont
  {D\"ur}}, \bibinfo {author} {\bibfnamefont {P.}~\bibnamefont {Sekatski}},\
  and\ \bibinfo {author} {\bibfnamefont {M.}~\bibnamefont {Skotiniotis}},\
  }\bibfield  {title} {\bibinfo {title} {Deterministic superreplication of
  one-parameter unitary transformations},\ }\href
  {https://doi.org/10.1103/PhysRevLett.114.120503} {\bibfield  {journal}
  {\bibinfo  {journal} {Phys. Rev. Lett.}\ }\textbf {\bibinfo {volume} {114}},\
  \bibinfo {pages} {120503} (\bibinfo {year} {2015})}\BibitemShut {NoStop}%
\bibitem [{\citenamefont {Chiribella}\ \emph {et~al.}(2015)\citenamefont
  {Chiribella}, \citenamefont {Yang},\ and\ \citenamefont
  {Huang}}]{GatecloningChiribella2015}%
  \BibitemOpen
  \bibfield  {author} {\bibinfo {author} {\bibfnamefont {G.}~\bibnamefont
  {Chiribella}}, \bibinfo {author} {\bibfnamefont {Y.}~\bibnamefont {Yang}},\
  and\ \bibinfo {author} {\bibfnamefont {C.}~\bibnamefont {Huang}},\ }\bibfield
   {title} {\bibinfo {title} {Universal superreplication of unitary gates},\
  }\href {https://doi.org/10.1103/PhysRevLett.114.120504} {\bibfield  {journal}
  {\bibinfo  {journal} {Phys. Rev. Lett.}\ }\textbf {\bibinfo {volume} {114}},\
  \bibinfo {pages} {120504} (\bibinfo {year} {2015})}\BibitemShut {NoStop}%
\end{thebibliography}%
\end{document}